\documentclass[12pt]{article}
\usepackage{amsmath}
\usepackage{amssymb}
\usepackage{graphicx}
\usepackage{epsfig}
\usepackage{cite}
\usepackage{url}
\usepackage[small]{caption}
\begin{document}
\baselineskip 0.6cm

\def\simgt{\mathrel{\lower2.5pt\vbox{\lineskip=0pt\baselineskip=0pt
           \hbox{$>$}\hbox{$\sim$}}}}
\def\simlt{\mathrel{\lower2.5pt\vbox{\lineskip=0pt\baselineskip=0pt
           \hbox{$<$}\hbox{$\sim$}}}}
\def\simprop{\mathrel{\lower3.0pt\vbox{\lineskip=1.0pt\baselineskip=0pt
             \hbox{$\propto$}\hbox{$\sim$}}}}
\def\bra#1{\langle #1 |}
\def\ket#1{| #1 \rangle}
\def\inner#1#2{\left< #1 | #2 \right>}

\begin{titlepage}

\begin{flushright}
UCB-PTH-13/09\\
\end{flushright}

\vskip 1.5cm

\begin{center}
{\Large \bf The Entropy of a Vacuum: What Does the Covariant Entropy Count?}

\vskip 0.7cm

{\large Yasunori Nomura and Sean J. Weinberg}

\vskip 0.4cm

{\it Berkeley Center for Theoretical Physics, Department of Physics,\\
 University of California, Berkeley, CA 94720, USA}

\vskip 0.1cm

{\it Theoretical Physics Group, Lawrence Berkeley National Laboratory,
 CA 94720, USA}

\vskip 0.8cm

\abstract{We argue that a unitary description of the formation and 
 evaporation of a black hole implies that the Bekenstein-Hawking entropy 
 is the ``entropy of a vacuum'':\ the logarithm of the number of possible 
 independent ways in which quantum field theory on a fixed classical 
 spacetime background can emerge in a full quantum theory of gravity. 
 In many cases, the covariant entropy counts this entropy---the degeneracy 
 of emergent quantum field theories in full quantum gravity---with the 
 entropy of particle excitations in each quantum field theory giving only 
 a tiny perturbation.  In the Rindler description of a (black hole) horizon, 
 the relevant vacuum degrees of freedom manifest themselves as an extra 
 hidden quantum number carried by the states representing the second 
 exterior region; this quantum number is invisible in the emergent 
 quantum field theory.  In a distant picture, these states arise as 
 exponentially degenerate ground and excited states of the intrinsically 
 quantum gravitational degrees of freedom on the stretched horizon. 
 The formation and evaporation of a black hole involve processes in 
 which the entropy of collapsing matter is transformed into that of 
 a vacuum and then to that of final-state Hawking radiation.  In the 
 intermediate stage of this evolution, entanglement between the vacuum 
 and (early) Hawking radiation develops, which is transferred to the 
 entanglement among final-state Hawking quanta through the evaporation 
 process.  The horizon is kept smooth throughout the evolution; in 
 particular, no firewall develops.  Similar considerations also apply 
 for cosmological horizons, for example for the horizon of a meta-stable 
 de~Sitter space.}

\end{center}
\end{titlepage}

\section{Introduction and Summary}
\label{sec:intro}

Despite much effort, there remains confusion about how a quantum theory 
of gravity works, especially in dynamical spacetime.  Much of this 
confusion arises from the lack of clear understanding of the relation 
between the classical description of gravity, as suggested by general 
relativity, and the structure/dynamics of the microscopic degrees of 
freedom from which the classical picture of spacetime is supposed to 
arise.  A major step toward such an understanding was the discovery of 
the Bekenstein-Hawking entropy~\cite{Bekenstein:1973ur}, which suggests 
that a black hole---despite its unique nature in general relativity---is 
somehow associated with ${\cal A}/4l_{\rm P}^2$ quantum degrees of 
freedom, where ${\cal A}$ is the area of the horizon and $l_{\rm P} 
\simeq 1.62 \times 10^{-35}~{\rm m}$ the Planck length.  A question, 
however, remains.  Where are these degrees of freedom?  In other words, 
what does this entropy count?

A naive interpretation of the Bekenstein-Hawking entropy as the 
entanglement entropy between the interior and exterior regions within 
the framework of a quantum field theory on fixed classical spacetime 
background leads to the fundamental loss of information, which contradicts 
the basic principles of quantum mechanics~\cite{Hawking:1976ra}.  To 
avoid this problem, a unitary description of the black hole formation 
and evaporation processes was put forward~\cite{'tHooft:1990fr}---when 
viewed from a distance, a {\it complete} description of these processes 
is obtained in terms of the degrees of freedom located on and outside 
the stretched horizon (a surface located about $l_{\rm P}$ proper 
distance away from the mathematical horizon), and the interior spacetime 
arises manifestly only after one adopts a different, though equivalent, 
infalling description~\cite{Susskind:1993if}.  This ``complementarity'' 
picture beautifully addresses some of the possible issues associated 
with the unitary description, in particular possible cloning 
of infalling quantum information into the interior and exterior 
regions~\cite{Hayden:2007cs}, and it can form a basis of a consistent 
quantum mechanical treatment of eternally inflating multiverse 
cosmology~\cite{Nomura:2011dt,Nomura:2011rb}.  Recently, however, 
it was argued that complementarity cannot be a consistent 
story~\cite{Almheiri:2012rt,Almheiri:2013hfa,Marolf:2013dba,Braunstein:2009my}; 
the argument essentially boiled down to the incompatibility of 
the uniqueness of the (infalling) vacuum and the distant unitary 
description~\cite{Bousso:2013wia,Avery:2012tf}.  We disagreed with 
this conclusion~\cite{Nomura:2012ex,Nomura:2013nya,Nomura:2013gna}. 
We argued that the apparent problem had arisen from an overly simplistic 
picture on how classical spacetime emerges in a full quantum theory 
of gravity.  In particular, we argued that there are exponentially 
many black hole vacuum states corresponding to a single semi-classical 
black hole, and that there can be a semi-classical world built on 
{\it each} of them, all of which are described by the same quantum 
field theory on a fixed spacetime background although they represent 
different quantum states at the full quantum gravity level.

The aim of this paper is to elaborate further on the picture described 
just above and to elevate it to general statements about the relation 
between full quantum gravity and emergent quantum field theories in 
classical spacetime backgrounds.  Our basic points about black hole 
physics can be summarized as follow.
\begin{itemize}
\item
The Bekenstein-Hawking entropy is the ``entropy of a vacuum'':\ the 
logarithm of the number of possible independent ways in which quantum 
field theory on classical near-horizon black hole spacetime can emerge 
in a full quantum theory of gravity.  In general, the entropy of a system 
is given by the logarithm of the number of states $N$ with a specified 
macroscopic property of the system, which is given by the product of 
the number of vacuum states, $e^{S_{\rm vac}}$, and the number of 
excited states that can be built on each of them, $e^{S_{\rm mat}}$:
\begin{equation}
  N = e^{S_{\rm vac}} \times e^{S_{\rm mat}}.
\label{eq:num-states}
\end{equation}
For a black hole (more precisely, a near-horizon region of a black hole 
that can be described by a near-horizon theory), the entropies of the 
vacuum and matter, which includes massless matter i.e.\ radiation, are
\begin{equation}
  S_{\rm vac} \approx \frac{\cal A}{4l_{\rm P}^2},
\qquad
  S_{\rm mat} \approx O\left( \frac{{\cal A}^n}{l_{\rm P}^{2n}} \right),
\label{eq:Svac-Smat}
\end{equation}
where $n < 1$, presumably $n \simeq 3/4$~\cite{'tHooft:1993gx}.  We 
therefore find $S_{\rm vac} \gg S_{\rm mat}$, and the black hole entropy 
is given by
\begin{equation}
  S = S_{\rm vac} + S_{\rm mat} \approx S_{\rm vac}.
\label{eq:S_BH}
\end{equation}
We emphasize that the entropy here is the {\it fine-grained} entropy, i.e.\ 
the logarithm of the number of possible independent quantum states in which 
the {\it entire} system, including the horizon degrees of freedom, can be. 
This is the origin of the Bekenstein-Hawking entropy.
\item
When the black hole is described in a distant reference frame, the 
exponential degeneracy of the field theory vacuum states, indicated by 
$S_{\rm vac}$, arises from (intrinsically quantum gravitational) degrees 
of freedom on the stretched horizon.  In particular, we postulate that 
\begin{itemize}
\item[(a)] the stretched horizon degrees of freedom can take exponentially 
many different configurations, labeled by $k = 1,\cdots,e^{\approx 
{\cal A}/4 l_{\rm P}^2}$, which are (approximately) degenerate in energy, 
and all of which can comprise a field theory vacuum;
\item[(b)]
there are infinitely many internally excited states {\it for each of} 
these configurations, and these excited states as well as the ground state 
(for each $k$) are entangled with the near exterior states $\ket{i}$ in 
a specific manner, determined by Boltzmann factors, for a black hole to 
be in a vacuum state.
\end{itemize}
By labeling internal excitations by the index $\tilde{\imath}$, the black 
hole vacuum states can then be written as%
\footnote{Here and below, the sum over $i$ implies the corresponding 
 sum over $\tilde{\imath}$ as well; for example, $\sum_{i=1}^\infty 
 \ket{i} \ket{\tilde{\imath}; k} = \ket{1} \ket{\tilde{1}; k} + \ket{2} 
 \ket{\tilde{2}; k} + \cdots$.}
\begin{equation}
  \ket{\psi_k} = \frac{1}{\sqrt{Z}} \sum_i 
    e^{-\frac{\beta}{2}E_i} \ket{i} \ket{\tilde{\imath}; k};
\qquad
  Z = \sum_i e^{-\beta E_i},
\label{eq:psi_k-intro}
\end{equation}
where $E_i$ is the energy of the state $\ket{i}$ measured in the asymptotic 
region, and $\beta$ the inverse Hawking temperature.
\item
The structure described above implies that the stretched horizon degrees 
of freedom provide the states necessary to compose (the $e^{\approx 
{\cal A}/4 l_{\rm P}^2}$ copies of) the second exterior region of the 
Rindler space, and hence the near-horizon eternal black hole geometry. 
In fact, the quantum mechanical structure of a collapse-formed black 
hole after the horizon is stabilized to a generic state is, at each 
instant of time, well approximated by that of an eternal black hole 
at the microscopic level.  In particular, the form of the states in 
Eq.~(\ref{eq:psi_k-intro}) allows us to define the mode operators acting 
on $\ket{\tilde{\imath}; k}$ (which can be interpreted to have arisen 
as collective excitations of the stretched horizon degrees of freedom) 
and hence the infalling mode operators {\it for each $k$}, following 
the standard Unruh-Israel prescription~\cite{Unruh:1976db}.  These 
infalling mode operators satisfy the algebra
\begin{equation}
  [a_{\sigma}^{(k)}, a_{\sigma'}^{(k')\dagger}] 
  = \delta_{\sigma\sigma'} \delta_{kk'},
\qquad
  [a_{\sigma}^{(k)}, a_{\sigma'}^{(k')}] 
  = [a_{\sigma}^{(k)\dagger}, a_{\sigma'}^{(k')\dagger}] 
  = 0,
\label{eq:gen-comm}
\end{equation}
where $\sigma$ represents a set of spacetime and internal quantum numbers, 
and $a_{\sigma}^{(k)}$ annihilates all the vacuum states, i.e.
\begin{equation}
  a_{\sigma}^{(k)} \ket{\psi_{k'}} = 0,
\label{eq:gen-vac}
\end{equation}
for all $\sigma, k, k'$.  A state in which matter exists in the interior 
of the black hole can be constructed by acting (a finite number of) 
$a_\sigma^{(k)\dagger}$'s on $\ket{\psi_k}$.
\item
The internal dynamics of the stretched horizon is such that the time 
evolution operator describing physics of an infalling object can be 
organized in a way that makes it manifest that the object smoothly passes 
through the horizon.  In particular, the Hamiltonian describing the 
infalling object can be written as
\begin{equation}
  H = \sum_{k=1}^{e^{\approx {\cal A}/4 l_{\rm P}^2}}\! 
    H_{\rm UI}\bigl( a_\sigma^{(k)}, a_\sigma^{(k)\dagger}; 
    c_{\bf p}, c_{\bf p}^\dagger \bigr)\, P_k,
\label{eq:H_infall}
\end{equation}
where $H_{\rm UI}(a_\sigma, a_\sigma^\dagger; c_{\bf p}, c_{\bf p}^\dagger)$ 
is the Hamiltonian in the standard Unruh-Israel description, with 
$a_\sigma$ and $c_{\bf p}$ being infalling mode operators and operators 
for far exterior modes, respectively, and $P_k$ is the projection 
operator defined by $P_k \ket{\tilde{\imath}; k'} = \delta_{kk'} 
\ket{\tilde{\imath}; k}$.  This implies that an infalling observer 
finds that the horizon is smooth with a probability of $1$.  When 
the observer interacts with the black hole state, which involves 
both the stretched horizon states and the near exterior states with 
which the stretched horizon states are strongly entangled, he/she 
``measures'' the black hole in the basis $\{ \ket{\psi_k} \}$, all 
of which lead to the same semi-classical physics predicted by general 
relativity.
\item
When viewed from a distance, unitarity of the black hole evolution is 
preserved in such a way that, at an intermediate stage of the evolution, 
the information about the initial collapsing matter is encoded in how 
the field theory vacuum on the black hole background is realized at the 
fundamental quantum gravity level, which will later be transformed into 
the state of final-state Hawking radiation.  The flow of information is 
thus schematically
\begin{equation}
  \mbox{collapsing matter} \longrightarrow 
  \begin{array}{c} \mbox{field theory vacuum} \\ 
    \mbox{(index }k\mbox{)}  \end{array} 
  \longrightarrow \mbox{Hawking radiation}.
\label{eq:inf-flow}
\end{equation}
This exchange of information between matter/radiation and a vacuum is 
a characteristic feature of the black hole formation and evaporation 
processes.  Note that physics outside the stretched horizon can still 
be completely local in the conventional sense throughout this process.%
\footnote{{\bf Note added:} More recently, this process has been analyzed 
 in detail in Ref.~\cite{Nomura:2014yka}.  In order for the process to 
 be local, e.g.\ to respect causality in spacetime, a (small) portion of 
 the information about the index $k$ must be regarded as being delocalized 
 into the whole zone region, $r \simlt 3M l_{\rm P}^2$, at the field 
 theory level.  In our notation here, this implies that a part of the 
 information about $k$ must be carried by the near exterior state $\ket{i}$ 
 in Eq.~(\ref{eq:psi_k-intro}), instead of $\ket{\tilde{\imath}; k}$. 
 Since the amount of information that needs to be delocalized to the 
 whole zone region is much smaller than ${\cal A}/4 l_{\rm P}^2$, 
 however, our discussions in this paper may persist essentially without 
 changes (except for points related to the issue described here). 
 For more complete and updated discussions on these points, see 
 Ref.~\cite{Nomura:2014yka}.}
\end{itemize}

The features of black hole physics summarized above lead to the following 
picture on the emergence of quantum field theories (built on classical 
spacetime backgrounds) in the full theory of quantum gravity.  Starting 
from the theory of the most fundamental quantum degrees of freedom, we 
can make a ``classical approximation'' only on certain degrees of freedom, 
corresponding to spacetime, keeping the full quantum nature for the 
rest of the degrees of freedom---this is what quantum field theory is. 
This classical approximation, as usual in such approximations, involves 
coarse-graining huge degrees of freedom.  As in many other systems, 
however, the entropy associated with these degrees of freedom---the 
entropy of a vacuum $S_{\rm vac}$---is still visible in the coarse-grained 
theory, i.e.\ quantum field theory, through thermodynamic considerations. 
A special feature of the black hole evaporation process in this respect 
is that the information contained in fine-grained degrees of freedom 
(i.e.\ constituents of spacetime) can get back to that in coarse-grained 
degrees of freedom (i.e.\ Hawking radiation), which does not happen 
in many systems.  Since our macroscopic world ultimately appears after 
many of the coarse-grained degrees of freedom, i.e.\ matter, are also 
classicalized, we may say that quantum field theories represent ``intermediate 
approximations'' in which a (major) part of the classicalization needed 
to go from the most fundamental theory to our classical world is taken 
into account explicitly.

Armed with the lessons we learned in our study of black hole physics, we 
also extend our considerations to more general cases.  We first consider 
a relatively straightforward application of the dynamics of the stretched 
horizon described above to a de~Sitter horizon.  We argue that, as 
in the black hole case, the stretched horizon degrees of freedom are 
organized into states labeled as $\ket{\tilde{\imath}; k}$, and that the 
de~Sitter vacuum states take the form in Eq.~(\ref{eq:psi_k-intro}), 
where $\ket{i}$ represents states in the interior of the stretched 
de~Sitter horizon.  The index $k$ runs over $1,\cdots,e^{\approx {\cal A}/4 
l_{\rm P}^2}$; here, ${\cal A} = 4\pi/H^2$ is the area of the (stretched) 
de~Sitter horizon, where $H$ is the Hubble parameter.  The same analysis 
as in the black hole case implies that a de~Sitter horizon is smooth:\ 
an object that hits the horizon can be thought of as going to space outside 
the horizon.  The information about the object that goes outside will be 
stored in the state constructed purely from the interior and the stretched 
horizon degrees of freedom.  Such information may thus be recovered 
later.  This recovery may not necessarily be in the form of Hawking 
radiation, if the system evolves, for example, into Minkowski space 
or another de~Sitter space with a smaller vacuum energy.

We also discuss implications of our observations for more general 
spacetimes in quantum gravity.  Here we adopt the picture advocated 
in Refs.~\cite{Nomura:2011rb,Nomura:2013nya} that the Hilbert space for 
quantum gravity can be organized in such a way that the system is viewed 
from a freely falling (local Lorentz) reference frame.  We argue that 
in general the fine-grained entropy of the system arises from both vacuum 
and matter/radiation contributions, and conjecture that it saturates the 
covariant entropy bound~\cite{Bousso:1999xy} if the degrees of freedom 
on the horizon (which may be located at spatial infinity as in Minkowski 
space) are included:
\begin{equation}
  S = S_{\rm vac} + S_{\rm mat} \approx \frac{\cal A}{4 l_{\rm P}^2}.
\label{eq:S-general}
\end{equation}
For the contribution from a horizon at which Planckian physics is important, 
such as the black hole or de~Sitter horizon, the vacuum contribution 
typically dominates:\ $S_{\rm vac} \gg S_{\rm mat}$.

The framework presented in this paper largely builds on the 
basic picture presented in Ref.~\cite{Nomura:2013gna} (and particularly 
emphasized in~\cite{Nomura}) that the number of degrees of freedom 
relevant to describe the black hole interior is a tiny fraction of 
the total number of degrees of freedom available for a black hole. 
The explicit realization of the idea, however, is different in 
this paper.  In Ref.~\cite{Nomura:2013gna}, it was considered, 
following~\cite{Verlinde:2013uja}, that a variety of black hole vacuum 
states $\ket{\psi_k}$ is allowed because of a freedom in the way the 
near horizon and stretched horizon modes are entangled; specifically, 
we considered a freedom in quantum mechanical phase factors appearing 
in the entangled states.  This approach, however, leads to the 
following issue.  When an arbitrary black hole state $\ket{\psi} 
= \sum_k c_k \ket{\psi_k}$ is considered, it in general does not 
lead to the thermal density matrix in the exterior region, implying 
that radiation emitted from the black hole does not have the Hawking 
spectrum.  (The population probability deviates by an $O(1)$ fraction 
for each energy level.)  The spectrum approximately looks like the 
Hawking form if a sufficient coarse-graining is performed in energy 
(or we may recover the exact Hawking form if we postulate particular 
dynamics that enforces a very specific form of entanglement between 
$\ket{\psi_k}$ and the environment), but it is still uncomfortable that 
the picture implied by semi-classical analyses receives such a major 
correction in the regime where we do not expect it.  In our framework 
presented here, the variety of the black hole vacuum states (index $k$) 
originates purely from the stretched horizon degrees of freedom.  This 
avoids the above issue of large deviations from the Hawking spectrum---the 
spectrum is exactly the Hawking form at each moment of emission (although 
there can be correlations between Hawking quanta emitted at different 
moments, needed to preserve unitarity)---and it allows us to construct 
interior quantum field theory operators explicitly by simply following 
the standard Unruh-Israel prescription.  The framework is also consistent 
with the criterion for a smooth horizon in Ref.~\cite{VanRaamsdonk:2010pw}:\ 
a smooth horizon requires a near-maximal entanglement in two sets of 
basis states, which was not the case in Ref.~\cite{Nomura:2013gna}.

The organization of the rest of the paper is as follows.  In 
Section~\ref{sec:BH-entropy}, we present our framework for describing 
black hole physics.  We motivate each of the hypotheses we introduce 
about the dynamics of the stretched horizon, and show that they allow 
us to reconstruct the interior spacetime consistently with unitarity 
of the evolution of the system.  We argue that the Bekenstein-Hawking 
entropy is the entropy of a vacuum.  In Section~\ref{sec:general}, we 
extend our considerations to more general cases.  We discuss how the 
same dynamics of the stretched horizon as in the black hole case applies 
to a de~Sitter horizon.  We also discuss the fine-grained entropy of 
more general spacetimes in quantum gravity and its relation to the 
covariant entropy bound.  In the appendix, we present the full Hilbert 
space needed to describe the evolution of a black hole.

While completing this paper, we received Ref.~\cite{Papadodimas:2013wnh} 
discussing the black hole interior in AdS/CFT, which seems to employ 
some similar ideas.

\section{Black Hole Entropy as the Entropy of a Vacuum}
\label{sec:BH-entropy}

In this section, we discuss our framework for black hole physics.

\subsection{A unitary description of black hole evolution}
\label{subsec:BH}

Suppose we describe the formation and evaporation of a black hole in a distant 
reference frame.  Following Refs.~\cite{'tHooft:1990fr,Susskind:1993if}, 
we postulate that there exists a unitary description which involves only 
the outside and the (stretched) horizon degrees of freedom of the black 
hole.  Suppose some matter that is not entangled with the rest of the 
system collapses into a black hole of mass $M_0$.  This process can be 
described as
\begin{equation}
  \ket{M_{\rm init}} \ket{E_{\rm init}} 
  \rightarrow \ket{\psi(M_0)} \ket{E},
\label{eq:BH-formation}
\end{equation}
where $\ket{M_{\rm init}}$ represents the initial state of matter, 
$\ket{\psi(M_0)}$ the state of the black hole shortly after the formation, 
and $\ket{E_{\rm init}}$ and $\ket{E}$ the states of the rest of the 
system at the respective moments.  (The meaning of the state of the black 
hole will become clearer later.)  Now, consider forming the black hole of 
the same mass (and angular momentum and charge) by collapsing matter in 
different initial states $\ket{M_{{\rm init},a}}$ ($a = 1,2,\cdots$). 
Unitarity then implies that the state of the black hole must also carry 
the index $a$:
\begin{equation}
  \ket{M_{{\rm init},a}} \ket{E_{\rm init}} 
  \rightarrow \ket{\psi_a(M_0)} \ket{E}.
\label{eq:BH-formation-a}
\end{equation}
Namely, there must be many different black hole quantum states that 
correspond to the same (semi-)classical black hole.

How many black hole microstates $n(M)$ are there for the black hole 
with a fixed mass $M$?  The validity of the generalized second law of 
thermodynamics suggests that it is given by the exponential of the 
Bekenstein-Hawking entropy
\begin{equation}
  n(M) = e^{\frac{{\cal A}}{4l_{\rm P}^2}} = e^{4\pi M^2 l_{\rm P}^2},
\label{eq:BH-k}
\end{equation}
where ${\cal A} = 16\pi M^2 l_{\rm P}^4$ is the area of the horizon; here, 
we have assumed for simplicity that the black hole under consideration 
is (well approximated by) a Schwarzschild black hole in 4-dimensional 
spacetime.%
\footnote{More precisely, $n(M)$ is the number of black hole states 
 with their masses in a range between $M$ and $M+\delta M$.  The precise 
 value of $\delta M$ is unimportant for our purposes because it only leads 
 to a logarithmic correction in the exponent of $n(M)$ (unless $\delta M$ 
 is chosen exponentially small), which we will ignore.  For definiteness, 
 one may take $\delta M$ to be of order the decay width of a black 
 hole to a lighter black hole and a Hawking quantum, $\delta M \sim 
 1/M l_{\rm P}^2$. \label{ft:M}}
Note that while the number of black hole states that can be directly 
formed by a single collapse, as in Eq.~(\ref{eq:BH-formation-a}), 
is much smaller than $e^{{\cal A}/4 l_{\rm P}^2}$ (presumably of 
order $e^{c {\cal A}^{3/4}/l_{\rm P}^{3/2}}$ where $c$ is an $O(1)$ 
coefficient~\cite{'tHooft:1993gx}), all the $e^{{\cal A}/4 l_{\rm P}^2}$ 
black hole states are expected to be realized for more complicated 
histories, for example by producing a larger black hole and then 
evaporating down.

The existence of exponentially many black hole microstates
\begin{equation}
  \ket{\psi_k(M)};
\qquad
  k = 1,\cdots,n(M)=e^{4\pi M^2 l_{\rm P}^2},
\label{eq:BH-microstates}
\end{equation}
allows a unitary description of the black hole formation and evaporation 
processes as viewed from a distant reference frame.  A crucial question is 
how this picture can be compatible with the implication of the equivalence 
principle that an infalling object does not feel anything special at 
the horizon, which seems to suggest that the black hole must be in the 
{\it unique} vacuum state from the viewpoint of the infalling object 
(after the scrambling time of order $t_{\rm sc} = M_0 l_{\rm P}^2 \ln(M_0 
l_{\rm P})$~\cite{Hayden:2007cs,Sekino:2008he} is passed since the formation, 
which we assume to be the case).  In particular, the Unruh-Israel description 
of the black hole seems to imply that it must be in a unique state in which 
the degrees of freedom in the ``two exterior regions'' are almost maximally 
entangled~\cite{Unruh:1976db}.  Our first goal then is to figure out what 
the relation is between the unitary description, in which the black hole 
state has the index $k$ as in Eq.~(\ref{eq:BH-microstates}), and the 
Unruh-Israel-type description which seems to imply the unique state.

\subsection{The Unruh-Israel description of a black hole}
\label{subsec:Unruh}

Recall that the Unruh-Israel description of a horizon is obtained by 
expanding a quantum field in two different sets of normal modes.  (Here 
we consider only a single bosonic quantum field for simplicity.  The 
extension to the other cases is straightforward.)  Let us denote the 
annihilation operators for the Minkowski (corresponding to the infalling) 
modes by $a_{\Omega,\xi}$ while those for the two exterior modes of the 
Rindler (distant) expansion by $b_{\omega,\xi}$ and $\tilde{b}_{\omega,\xi}$, 
respectively.  Here, $\xi$ collectively represents the quantum numbers 
associated with the directions parallel to the horizon, e.g.\ the two 
angular momentum quantum numbers $l$ and $m$ for a spherical horizon, 
and $\Omega$ and $\omega$ the frequencies.  The two sets of modes are 
related by a Bogoliubov transformation
\begin{equation}
  a_{\Omega,\xi} = \sum_\omega \bigl( \alpha_{\Omega,\omega} b_{\omega,\xi} 
    + \gamma_{\Omega,\omega} b_{\omega,\xi}^\dagger 
    + \zeta_{\Omega,\omega} \tilde{b}_{\omega,\xi} 
    + \eta_{\Omega,\omega} \tilde{b}_{\omega,\xi}^\dagger \bigr),
\label{eq:modes-rel}
\end{equation}
where $\alpha_{\Omega,\omega}$, $\gamma_{\Omega,\omega}$, 
$\zeta_{\Omega,\omega}$, and $\eta_{\Omega,\omega}$ are coefficients.

The Minkowski/infalling vacuum state $\ket{\psi}$ is defined by
\begin{equation}
  \forall \Omega, \xi, \quad
  a_{\Omega,\xi} \ket{\psi} = 0,
\label{eq:inf_vac}
\end{equation}
which is represented in the Rindler/distant frame by
\begin{equation}
  \ket{\psi} \propto \prod_{\omega,\xi} \exp\Bigl( 
    e^{-\frac{\beta \omega}{2}} b_{\omega,\xi}^\dagger 
    \tilde{b}_{\omega,\xi}^\dagger \Bigr) \ket{\emptyset},
\label{eq:inf_vac-Rind}
\end{equation}
where $\ket{\emptyset}$ is the Rindler vacuum and $\beta$ the inverse 
temperature.  The temperature $1/\beta$ may in general depend on the mode 
$(\omega,\xi)$, but this dependence is essentially absent when $\omega$ 
is defined in the asymptotic region, since the temperature and frequencies 
redshift in the same way.  (In the true Rindler space, this requires 
the introduction of an infrared cutoff.  In the black hole case, the 
Unruh-Israel description is valid only for a region close to the horizon, 
and the local Hawking temperature and mode frequencies at the outer edge 
of this region are scaled from their respective asymptotic values by 
the same $O(1)$ factor.)  We thus drop the possible dependence of $\beta$ 
on $(\omega,\xi)$ below, assuming that $\omega$ represents the frequency 
measured in the asymptotic region.

We now decompose $\ket{\emptyset}$ as
\begin{equation}
  \ket{\emptyset} = \ket{0} \ket{\tilde{0}},
\label{eq:0-decomp}
\end{equation}
where $\ket{0}$ and $\ket{\tilde{0}}$ are the vacuum states in 
the two exterior regions on which $b_{\omega,\xi}^\dagger$'s and 
$\tilde{b}_{\omega,\xi}^\dagger$'s act, respectively.  By tracing 
out the exterior states in one side (called Region~III), i.e.\ 
$\ket{\tilde{0}}$, $\tilde{b}_{\omega,\xi}^\dagger \ket{\tilde{0}}$, 
$\frac{1}{\sqrt{2^{\delta_{\omega\omega'} \delta_{\xi\xi'}}}} 
\tilde{b}_{\omega,\xi}^\dagger \tilde{b}_{\omega',\xi'}^\dagger 
\ket{\tilde{0}}$, $\cdots$, we obtain the density matrix in the other 
exterior region (Region~I):
\begin{equation}
  \rho_{\rm I} = \underset{\rm Region~III}{\rm Tr} \ket{\psi}\bra{\psi} 
  = \frac{1}{Z} \sum_i e^{-\beta E_i} \ket{i} \bra{i},
\label{eq:rho}
\end{equation}
where $Z = \sum_i e^{-\beta E_i}$.  Here, $\ket{i}$ represents a state 
in Region~I, specified by the number of excitations $n_{\omega,\xi}$ in 
each mode $(\omega,\xi)$:
\begin{equation}
  \ket{i} = \left( \prod_{\omega,\xi} \frac{1}{\sqrt{n_{\omega,\xi}!}} 
    (b_{\omega,\xi}^\dagger)^{n_{\omega,\xi}} \right) \ket{0},
\label{eq:ket-i}
\end{equation}
while $E_i = \sum_{\omega,\xi} n_{\omega,\xi} \omega$ is the energy 
of the state $\ket{i}$.  We identify Region~I to represent the side 
exterior to the Schwarzschild horizon, which leads to $\beta = 1/T_H 
= 8\pi M l_{\rm P}^2$, where $T_H$ is the Hawking temperature.  Note that 
the Unruh-Israel state provides a description only of a spacetime region 
close to the horizon, e.g.\ $r \simlt 3M l_{\rm P}^2$, so to describe 
the entire region outside the Schwarzschild horizon we need to consider 
quantum states describing the far region, e.g.\ $r \simgt 3M l_{\rm P}^2$, 
in addition to this state.  (The validity of this division of the entire 
system into two subsystems is granted by locality, which we assume to 
be preserved outside the stretched horizon at length scales larger than 
the fundamental, or string, length $l_*$.)  On the other hand, the interior 
region of the Schwarzschild horizon is described (fully) by quantum 
field theory built on $\ket{\psi}$ by acting $a_{\Omega,\xi}^\dagger$ 
operators (after an appropriate modification from a planer to the 
spherical horizon is made).

The von~Neumann entropy of the density matrix $\rho_{\rm I}$ in 
Eq.~(\ref{eq:rho}) is given by
\begin{equation}
  S_{\rm I} = - {\rm Tr} (\rho_{\rm I} \ln\rho_{\rm I}) 
  = \frac{{\cal A}}{4 l_{\rm P}^2} \left\{ 1 + 
    O\left( \frac{l_{\rm P}^{2n}}{{\cal A}^n};\, n>0 \right) \right\}
  \approx \frac{{\cal A}}{4 l_{\rm P}^2},
\label{eq:von_Neumann}
\end{equation}
where ${\cal A}$ is the area of the horizon~\cite{Sorkin}, implying that 
the number of terms in the last expression of Eq.~(\ref{eq:rho}) having 
unsuppressed coefficients is (effectively) $e^{\approx {\cal A}/4 
l_{\rm P}^2}$.  Here and below, the approximate symbol indicates that 
the expression is valid at the leading order in expansion in inverse 
powers of ${\cal A}/l_{\rm P}^2$.  Note that to obtain the coefficient 
of $1/4$ in Eq.~(\ref{eq:von_Neumann}), one needs to include the effect 
of the counterterm renormalizing Newton's constant.  This implies that 
the number of $1/4$ is obtained only after we include all the ultraviolet 
(including trans-Planckian) states in $\ket{i}$ in Eq.~(\ref{eq:rho}), 
which is implicitly done through the counterterm.

How should we interpret the Unruh-Israel result described above? 
The conventional interpretation is that it describes the {\it unique} 
infalling vacuum state in which a black hole must be at late times, 
specifically after the scrambling time.  Our interpretation is different---we 
consider that a black hole at late times consists, as suggested by unitarity, 
of exponentially many infalling vacuum states $\ket{\psi_k}$, and that 
the Unruh-Israel description arises as an emergent effective quantum 
field theory in {\it each} of these vacuum states, which is responsible 
for describing an object falling into the horizon.  As we will see in 
Section~\ref{subsec:infalling}, this structure allows us to avoid the 
arguments for firewalls and to keep the horizon smooth consistently with 
the unitarity of the black hole formation and evaporation processes. 
We will now discuss this picture in more detail.

\subsection{The entropy of a vacuum and emergent quantum field theories}
\label{subsec:ent_vac}

Our starting point is to adopt a set of hypotheses that we consider 
natural from the viewpoint of a distant reference frame:
\begin{itemize}
\item[(i)]
The formation and evaporation of a black hole are unitary processes. 
This implies that there are exponentially many black hole vacuum states 
$\ket{\psi_k}$.
\item[(ii)]
The number of black hole states $\ket{\psi_k}$ for a fixed mass $M$ is 
$n(M) = e^{\approx {\cal A}/4 l_{\rm P}^2}$, where ${\cal A} = 16\pi 
M^2 l_{\rm P}^4$.  This is motivated by the success of the generalized 
second law of thermodynamics.
\item[(iii)]
For any black hole state $\ket{\psi_k}$, the region near and outside the 
stretched horizon---which we assume to be well described by local quantum 
field theory---is given by the mixed, thermal state as in Eq.~(\ref{eq:rho}):
\begin{equation}
  \rho_{\rm ext} = \frac{1}{Z} \sum_i e^{-\beta E_i} \ket{i} \bra{i}.
\label{eq:rho_ext}
\end{equation}
Here, $\ket{i}$ represents the states near and {\it outside} the stretched 
horizon; in particular, it does not include the (Planckian) degrees of 
freedom associated with the stretched horizon.
\item[(iv)]
The state in Eq.~(\ref{eq:rho_ext}) is purified by the (intrinsically 
quantum gravitational) degrees of freedom located on the stretched horizon; 
namely, the stretched horizon degrees of freedom play the role of the 
second exterior region in the Unruh-Israel description~\cite{Nomura:2013gna}. 
Item~(ii) above then implies that there are $e^{\approx {\cal A}/4 
l_{\rm P}^2}$ different ways in which $\rho_{\rm ext}$ is purified 
by the stretched horizon states:
\begin{equation}
  \rho_{\rm ext} \rightarrow \ket{\psi_k},
\label{eq:purify}
\end{equation}
where $k = 1,\cdots,e^{\approx {\cal A}/4 l_{\rm P}^2}$.  Since the 
states $\ket{i}$ that have unsuppressed Boltzmann coefficients in 
Eq.~(\ref{eq:rho_ext}) represent modes localized near but outside the 
stretched horizon, the black hole state $\ket{\psi_k}$ must be thought 
of as representing the states of the stretched horizon degrees of freedom 
{\it as well as} the exterior modes represented by the $\ket{i}$'s, which 
are highly entangled with each other.  (Note that black hole states 
$\ket{\psi_k}$ can be further entangled with states representing the 
rest of the system, in which case the state of the black hole can 
only be represented by a density matrix in the space spanned by the 
$\ket{\psi_k}$'s; see Section~\ref{subsec:distant} for further discussion.)
\end{itemize}

We now postulate the following structure for the stretched horizon states. 
As suggested by the existence of exponentially many black hole states, 
Eq.~(\ref{eq:BH-microstates}), we consider that the stretched horizon 
degrees of freedom can take exponentially many different configurations 
which are (approximately) degenerate in energy.  We label these 
configurations by the index $k$, which runs over
\begin{equation}
  k = 1, \cdots, n(M) = e^{\approx \frac{{\cal A}}{4 l_{\rm P}^2}},
\label{eq:k}
\end{equation}
for a fixed black hole mass $M$.  (More precisely, the stretched 
horizon degrees of freedom can take $e^{\approx {\cal A}/4 l_{\rm P}^2}$ 
configurations in the energy range between $M$ and $M + \delta M$; see 
footnote~\ref{ft:M}.)  We consider that there are (an infinite number 
of) internally excited states for each of these configurations, and we 
label these excited as well as the ground states by $\tilde{\imath}$.  The 
stretched horizon states can then be denoted as $\ket{\tilde{\imath}; k}$, 
which form an orthonormal set
\begin{equation}
  \inner{\tilde{\imath}; k}{\tilde{\imath}'; k'} 
  = \delta_{\tilde{\imath}\tilde{\imath}'} \delta_{kk'}.
\label{eq:stretched}
\end{equation}
Motivated by the Unruh-Israel description, we consider that the ground 
and excited states for each $k$ are entangled with the near exterior 
states $\ket{i}$ in a specific manner, with the coefficients determined 
by Boltzmann factors.  In particular, we assume that the black hole 
vacuum states take the specifically entangled form
\begin{equation}
  \ket{\psi_k(M)} = \frac{1}{\sqrt{Z}} \sum_i 
    e^{-\frac{\beta}{2}E_i} \ket{i} \ket{\tilde{\imath}; k}.
\label{eq:phi_k}
\end{equation}
This structure satisfies all the requirements in (i)~--~(iv) above. 
In particular, upon integrating out the stretched horizon states 
$\ket{\tilde{\imath}; k}$, we find that the reduced density matrix for 
the exterior states takes the form of Eq.~(\ref{eq:rho_ext}) for any of 
the states $\ket{\psi_k(M)}$.  Based on a simple field theory estimate, 
we expect $S_{\rm ext} = -{\rm Tr} (\rho_{\rm ext} \ln\rho_{\rm ext}) 
\approx \gamma {\cal A}/l_{\rm P}^2$, where $\gamma \approx O(1)$. 
This implies that the number of terms in the right-hand side 
of Eq.~(\ref{eq:phi_k}) whose coefficients are nonnegligible 
is $e^{\approx \gamma {\cal A}/l_{\rm P}^2}$.%
\footnote{In the present picture, the quantity $S_{\rm ext}$ ($\approx 
 \gamma {\cal A}/l_{\rm P}^2$) is not related with the number of 
 black hole states $n(M)$ in Eq.~(\ref{eq:k}), so we generally expect 
 $\gamma \neq 1/4$.  In fact, $\gamma$ will be sensitive to the 
 precise division of the degrees of freedom into the stretched horizon 
 and near exterior modes, which is somewhat arbitrary.  Note that 
 the quantity $S_{\rm ext}$ here is different from $S_{\rm I}$ in 
 Eq.~(\ref{eq:von_Neumann}) in which the ultraviolet modes (corresponding 
 to the stretched horizon modes here) are included implicitly in 
 $\ket{i}$.  In our picture, $S_{\rm I}$ counts the {\it total} 
 number of effective degrees of freedom existing in one side of 
 the {\it mathematical} Schwarzschild horizon (which we view as 
 the only physical degrees of freedom), thus giving $S_{\rm I} \approx 
 \ln {\rm dim} \{ \ket{\psi_k(M)} \} \approx {\cal A}/4 l_{\rm P}^2$. 
 In other words, calculations such as in Ref.~\cite{Sorkin}---after 
 including the counterterm---count $\ln n(M)$ in the language here 
 (by considering the fictitious ``vacuum state'' $\sum_k \ket{\psi_k} 
 \ket{\tilde{\psi}_k}$).}

We now argue that we can build an effective quantum field theory 
describing the interior and near exterior regions of the horizon on 
each of the black hole vacuum states $\ket{\psi_k}$, and that all 
these quantum field theories are isomorphic with each other, taking the 
form identical to the Unruh-Israel description of the black hole.  (In 
Section~\ref{subsec:infalling}, we will argue that this quantum field 
theory is the one responsible for describing the fate of an infalling 
object.)  Let us consider states built on a specific black hole vacuum 
state $\ket{\psi_k}$.  Since the states $\ket{i}$ are specified by a set 
of occupation numbers $n_{\omega,\xi}$ for all the modes $(\omega,\xi)$, 
we write them as
\begin{equation}
  \ket{i} \rightarrow \ket{\{ n_{\omega,\xi} \}}.
\label{eq:i_n}
\end{equation}
Similarly, we can write the stretched horizon states as
\begin{equation}
  \ket{\tilde{\imath}; k} \rightarrow \ket{\{ \tilde{n}_{\omega,\xi} \}; k},
\label{eq:itilde_n}
\end{equation}
where $\tilde{n}_{\omega,\xi}$'s are the occupation numbers of the 
$(\omega,\xi)$ modes {\it of $\ket{i}$}, not $\ket{\tilde{\imath}; k}$, 
to which $\ket{\tilde{\imath}; k}$ is coupled {\it in the vacuum 
state $\ket{\psi_k}$} in Eq.~(\ref{eq:phi_k}).  (Note that $\{ 
\tilde{n}_{\omega,\xi} \}$ here is simply used to label states 
$\ket{\tilde{\imath}; k}$ through Eq.~(\ref{eq:phi_k}), whose meaning 
is still the occupation numbers for the exterior states $\ket{i}$. 
We will, however, see later that it can be understood as the occupation 
numbers for some ``quasi-particles'' represented by the stretched 
horizon states $\ket{\tilde{\imath}; k}$.)  A general black hole---not 
necessarily vacuum---state $\ket{\phi_k}$ obtained by exciting 
$\ket{\psi_k}$ can then be written as
\begin{equation}
  \ket{\phi_k} = \frac{1}{\sqrt{Z_\phi}} 
    \sum_{\{ n_{\omega,\xi} \}, \{ \tilde{n}_{\omega,\xi} \}} 
    f_{\{ n_{\omega,\xi} \}, \{ \tilde{n}_{\omega,\xi} \}}\, 
    \ket{\{ n_{\omega,\xi} \}} \ket{\{ \tilde{n}_{\omega,\xi} \}; k},
\label{eq:phi-gen}
\end{equation}
where $Z_\phi = \sum_{\{ n_{\omega,\xi} \}, \{ \tilde{n}_{\omega,\xi} \}} 
|f_{\{ n_{\omega,\xi} \}, \{ \tilde{n}_{\omega,\xi} \}}|^2$.  The vacuum 
state $\ket{\psi_k}$ is a special case in which
\begin{equation}
  f_{\{ n_{\omega,\xi} \}, \{ \tilde{n}_{\omega,\xi} \}} 
  = e^{-\frac{\beta}{2} E_{\{ n_{\omega,\xi} \}}} 
    \delta_{\{ n_{\omega,\xi} \} \{ \tilde{n}_{\omega,\xi} \}},
\label{eq:vac_k}
\end{equation}
where $\delta_{\{ n_{\omega,\xi} \} \{ \tilde{n}_{\omega,\xi} \}} \equiv 
\prod_{\omega,\xi} \delta_{n_{\omega,\xi} \tilde{n}_{\omega,\xi}}$.

Suppose that we define operators $\tilde{b}_{\omega,\xi}^{(k)}$ and 
$\tilde{b}_{\omega,\xi}^{(k)\dagger}$ acting on the stretched horizon 
degrees of freedom by
\begin{align}
  \tilde{b}_{\omega,\xi}^{(k)} \ket{\{ \tilde{n}_{\omega',\xi'} \}; k'} 
  &= \delta_{kk'} \sqrt{\tilde{n}_{\omega,\xi}}\, 
      \ket{\{ \tilde{n}_{\omega',\xi'} - 
      \delta_{\omega\omega'}\delta_{\xi\xi'} \}; k},
\label{eq:tilde-b}\\
  \tilde{b}_{\omega,\xi}^{(k)\dagger} \ket{\{ \tilde{n}_{\omega',\xi'} \}; k'} 
  &= \delta_{kk'} \sqrt{\tilde{n}_{\omega,\xi} + 1}\, 
      \ket{\{ \tilde{n}_{\omega',\xi'} + 
      \delta_{\omega\omega'}\delta_{\xi\xi'} \}; k}.
\label{eq:tilde-b-dag}
\end{align}
The vacuum state $\ket{\psi_k}$ can then be written as in 
Eq.~(\ref{eq:phi_k}):
\begin{equation}
  \ket{\psi_k} = \frac{1}{\sqrt{Z}} \sum_{i =  \{ n_{\omega,\xi} \}} 
    e^{-\frac{\beta E_i}{2}} \ket{i} \ket{\tilde{\imath}; k},
\label{eq:phi_k-2}
\end{equation}
with $\ket{\tilde{\imath}; k}$'s now given by
\begin{equation}
  \ket{\tilde{\imath}; k} = \left( \prod_{\omega,\xi} 
    \frac{1}{\sqrt{n_{\omega,\xi}!}} 
    (\tilde{b}_{\omega,\xi}^{(k)\dagger})^{n_{\omega,\xi}} \right) 
    \ket{\tilde{0}; k}.
\label{eq:ket-tilde-i}
\end{equation}
Here, $\ket{\tilde{0}; k}$ is the ground state of the $k$-th configuration 
of the stretched horizon degrees of freedom, satisfying
\begin{equation}
  \forall \omega, \xi, \quad
  \tilde{b}_{\omega,\xi}^{(k)} \ket{\tilde{0}; k} = 0.
\label{eq:tilde-0}
\end{equation}
The commutation relations between operators $\tilde{b}_{\omega,\xi}^{(k)}$ 
and $\tilde{b}_{\omega,\xi}^{(k)\dagger}$ are obtained from 
Eqs.~(\ref{eq:tilde-b},~\ref{eq:tilde-b-dag}) as
\begin{equation}
  [\tilde{b}_{\omega,\xi}^{(k)}, \tilde{b}_{\omega',\xi'}^{(k')\dagger}] 
  = \delta_{\omega\omega'} \delta_{\xi\xi'} \delta_{kk'},
\qquad
  [\tilde{b}_{\omega,\xi}^{(k)}, \tilde{b}_{\omega',\xi'}^{(k')}] 
  = [\tilde{b}_{\omega,\xi}^{(k)\dagger}, 
      \tilde{b}_{\omega',\xi'}^{(k')\dagger}] = 0.
\label{eq:b_k-comm}
\end{equation}
This implies that we can interpret $\tilde{b}_{\omega,\xi}^{(k)}$ and 
$\tilde{b}_{\omega,\xi}^{(k)\dagger}$ as the annihilation and creation 
operators for ``quasi-particle'' quanta with negative energy $-\omega$, 
which arise as (collective) excitation modes of the stretched horizon 
degrees of freedom.  As becomes clear below, these are precisely the 
Hawking partner modes that can be excited in the $k$-th vacuum state 
$\ket{\psi_k}$.

For a fixed $k$, the form of Eqs.~(\ref{eq:phi_k-2}~--~\ref{eq:b_k-comm}) 
is identical to that in the standard Unruh-Israel description.  In view 
of this, we define the infalling mode operators associated with the $k$-th 
vacuum state $\ket{\psi_k}$ by
\begin{equation}
  a_{\Omega,\xi}^{(k)} 
  = \sum_\omega \bigl( \alpha_{\Omega,\omega} b_{\omega,\xi} P_k 
    + \gamma_{\Omega,\omega} b_{\omega,\xi}^\dagger P_k 
    + \zeta_{\Omega,\omega} \tilde{b}_{\omega,\xi}^{(k)} 
    + \eta_{\Omega,\omega} \tilde{b}_{\omega,\xi}^{(k)\dagger} \bigr),
\label{eq:a_k}
\end{equation}
where $P_k$ is a projection operator acting on the stretched horizon 
states as $P_k \ket{\{ \tilde{n}_{\omega,\xi} \}; k'} = \delta_{kk'} 
\ket{\{ \tilde{n}_{\omega,\xi} \}; k}$, and $\alpha_{\Omega,\omega}$, 
$\gamma_{\Omega,\omega}$, $\zeta_{\Omega,\omega}$, and $\eta_{\Omega,\omega}$ 
are the same coefficients as in Eq.~(\ref{eq:modes-rel}). 
Note that $a_{\Omega,\xi}^{(k)}$ involves projection on $k$:\ 
$a_{\Omega,\xi}^{(k)} = a_{\Omega,\xi}^{(k)} P_k$ because 
of Eqs.~(\ref{eq:tilde-b},~\ref{eq:tilde-b-dag}).  These 
operators satisfy the algebra for creation/annihilation operators
\begin{equation}
  [a_{\Omega,\xi}^{(k)}, a_{\Omega',\xi'}^{(k')\dagger}] 
  = \delta_{\Omega\Omega'} \delta_{\xi\xi'} \delta_{kk'},
\qquad
  [a_{\Omega,\xi}^{(k)}, a_{\Omega',\xi'}^{(k')}] 
  = [a_{\Omega,\xi}^{(k)\dagger}, a_{\Omega',\xi'}^{(k')\dagger}] 
  = 0,
\label{eq:a_k-comm}
\end{equation}
and their actions on the vacuum states are given by
\begin{equation}
  \forall \Omega, \xi, k, \quad
  a_{\Omega,\xi}^{(k)} \ket{\psi_k} = 0,
\label{eq:inf_vac_k}
\end{equation}
(and $a_{\Omega,\xi}^{(k)} \ket{\psi_{k'}} = a_{\Omega,\xi}^{(k)\dagger} 
\ket{\psi_{k'}} = 0$ for $k \neq k'$).  We can therefore construct quantum 
states in an infalling reference frame by acting $a_{\Omega,\xi}^{(k)\dagger}$ 
operators on the vacuum state $\ket{\psi_k}$ {\it for each $k$}.

How many quantum states can we build on each $\ket{\psi_k}$?  By acting 
(a finite number of) $a_{\Omega,\xi}^{(k)\dagger}$'s on $\ket{\psi_k}$, 
one can construct a state in which matter exists in the interior of the 
black hole of mass $M$, when viewed from an infalling observer.%
\footnote{Strictly speaking, operating $a_{\Omega,\xi}^{(k)\dagger}$'s 
 on $\ket{\psi_k}$ changes the mass of the black hole.  This, however, 
 can be compensated by adjusting the mass associated with the singularity 
 at the center.  The argument below persists with this adjustment.}
We expect that the number of such states is of order $e^{\approx 
{\cal A}^n / l_{\rm P}^{2n}}$ with $n < 1$, presumably $n \simeq 
3/4$~\cite{'tHooft:1993gx}.  The number of all the black hole states 
of mass $M$ is therefore
\begin{equation}
  \Bigl[ n(M) = e^{\approx \frac{\cal A}{4l_{\rm P}^2}} \Bigr] 
    \times e^{\approx c \frac{{\cal A}^n}{l_{\rm P}^{2n}}}
  = e^{\approx \frac{\cal A}{4l_{\rm P}^2}},
\label{eq:n-classical}
\end{equation}
which is consistent with the holographic/covariant entropy 
bound~\cite{'tHooft:1993gx,Susskind:1994vu,Bousso:1999xy}.  In 
particular, Eq.~(\ref{eq:n-classical}) implies that the black hole 
(or covariant) entropy is saturated by {\it the entropy of a vacuum}
\begin{equation}
  S_{\rm vac} = \ln \{ n(M) \} \approx \frac{\cal A}{4l_{\rm P}^2},
\label{eq:S_vac}
\end{equation}
at the leading order in $l_{\rm P}^2/{\cal A}$, and that the entropy 
from usual matter (and radiation), $S_{\rm mat} \approx {\cal A}^n / 
l_{\rm P}^{2n}$ ($n < 1$), gives only a sub-leading contribution. 
(This statement applies even if we include all the near-horizon---not 
necessarily interior---excitations obtained by acting the 
$a_{\Omega,\xi}^{(k)\dagger}$'s.)  As will be seen more explicitly 
in Section~\ref{subsec:infalling}, $e^{S_{\rm vac}}$ represents the 
number of possible independent ways in which quantum field theory on 
a fixed classical spacetime background, which allows for the description 
of the black hole interior, can emerge in a full quantum theory of gravity. 
The black hole entropy mostly counts the logarithm of this number!

\subsection{The formation and evaporation of a black hole---the distant view}
\label{subsec:distant}

We now discuss how the black hole formation and evaporation processes 
are described from the point of view of a distant reference frame.  Here 
we focus on basic aspects of this description.  A detailed discussion 
on the structure of the Hilbert space, relevant to describe the evolution 
of a black hole, is given in the Appendix.

Consider collapsing matter represented by state $\ket{M_{\rm init}}$ that 
has not been entangled with the rest of the system $\ket{r_{\rm init}}$. 
The formation of a black hole (of the initial mass $M_0$) is then described 
as the following evolution of the entire system:
\begin{equation}
  \ket{\Psi} = \ket{M_{\rm init}} \ket{r_{\rm init}} 
  \rightarrow \left( \sum_{k=1}^{e^{\approx {\cal A}(t)/4 l_{\rm P}^2}}\!\! 
    c_k(t)\, \bigl|\psi_k(M(t))\bigr\rangle \right) \bigl|r(t)\bigr\rangle,
\label{eq:young}
\end{equation}
where ${\cal A}(t) = 16\pi M^2(t) l_{\rm P}^4$ is the area of the horizon 
at time $t$, and $M(t) = M_0$ at the time of the formation.%
\footnote{To be more precise, states well after the formation involve 
 superpositions of black hole states with different $M$'s, reflecting 
 the probabilistic nature of Hawking emission.  We ignore this effect, 
 as well as a possible spread of the initial black hole mass caused, 
 e.g., by quantum effects associated with the collapse, since they do 
 not affect our argument.}
Strictly speaking, the black hole shortly after the formation is not 
yet in a vacuum state represented by (a superposition of) $\ket{\psi_k}$'s; 
instead, it is in a more general state represented by $\ket{\phi_k}$'s in 
Eq.~(\ref{eq:phi-gen}).  After the scrambling time of order $t_{\rm sc} 
= M_0 l_{\rm P}^2 \ln(M_0 l_{\rm P})$, however, the black hole state 
takes the form shown in the biggest parentheses in the last expression, 
with $c_k$'s expected to take generic values
\begin{equation}
  |c_k(t)|^2 \sim O\biggl( e^{\approx -\frac{{\cal A}(t)}{4 
    l_{\rm P}^2}} \biggr).
\label{eq:generic}
\end{equation}
At this early stage in the evolution of the black hole, the state of the 
entire system is well approximated by the expression in Eq.~(\ref{eq:young}). 
In particular, entanglement between the black hole and the rest of the 
system may still be neglected (for more precise discussion, see below).

As time passes, however, the black hole becomes more and more entangled 
with the rest of the system in the sense that the ratio of the entanglement 
entropy between the black hole and the rest, $S_{\rm ent}(t)$, to the 
Bekenstein-Hawking entropy, $S_{\rm BH}(t) = 4\pi M(t)^2 l_{\rm P}^2$, 
keeps growing, which saturates the maximum value $S_{\rm ent}(t) 
/ S_{\rm BH}(t) = 1$ after the Page time $t_{\rm Page} \sim M_0^3 
l_{\rm P}^4$~\cite{Page:1993wv}.  Therefore, the state of the 
system at late times must be written more explicitly 
as~\cite{Nomura:2012ex,Nomura:2013nya}
\begin{equation}
  \ket{\Psi(t)} = \sum_{k=1}^{e^{\approx {\cal A}(t)/4 l_{\rm P}^2}}\!\! 
    d_k(t)\, \bigl|\psi_k(M(t))\bigr\rangle\, \bigl|r_k(t)\bigr\rangle,
\label{eq:old}
\end{equation}
where $\ket{r_k}$'s represent states of the subsystem complement to the 
black hole, i.e.\ those for the region $r \simgt 3M l_{\rm P}^2$, which 
include states of the Hawking radiation emitted earlier.  In other words, 
at these late times the logarithm of the dimension of space spanned by 
the $\ket{r_k}$'s is of order $S_{\rm BH}$ (and equal to $S_{\rm BH}$ 
after the Page time), while at much earlier times it is negligible compared 
with $S_{\rm BH}$.  The state at early times, therefore, can be well 
approximated by the expression in Eq.~(\ref{eq:young}) for the purpose 
of discussing internal properties of the black hole.

As the black hole evaporation progresses, the information contained 
in a set of coefficients $\{ d_k(t) \}$ is gradually transferred into 
that in states $\ket{r_{k'}(t')}$ with $t' > t$, specifically in the 
correlations of Hawking quanta emitted at different moments.  The 
dynamics of this process is governed by the interaction Hamiltonian 
coupling the stretched horizon and exterior degrees of freedom, whose 
form is determined by the intrinsically quantum gravitational, Planckian 
physics.  We will be agnostic about the precise form of this Hamiltonian.%
\footnote{{\bf Note added:} At the level of semi-classical field 
 theory, the information transfer is viewed as occurring through 
 a part of the information about the vacuum state which is delocalized 
 in the whole zone region.  For further discussions on this point, 
 see Ref.~\cite{Nomura:2014yka}.}
After the black hole has completely evaporated, the state of the 
system becomes that of the final-state Hawking quanta (and matter 
that did not collapse). At some late time, we denote it by $\ket{\Psi} 
= \ket{r_{\rm fin}}$.

At each moment in the evolution, the state of the black hole can be given 
by a density matrix in the space spanned by the $\ket{\psi_k}$'s, obtained 
by integrating out the rest of the system:
\begin{equation}
  \rho_{\rm BH} = \sum_{k,l} f_{kl} \ket{\psi_k} \bra{\psi_l},
\label{eq:BH-mixed}
\end{equation}
where $f_{kl}$ is a positive semi-definite Hermitian matrix with 
$\sum_k f_{kk} = 1$.  (At early stages in which the entanglement between 
the black hole and the rest is neglected, $f_{kl}$ takes the form 
$\simeq c_k c_l^*$; in other words, the black hole can be represented 
by a pure state $\ket{\psi_{\rm BH}} = \sum_k c_k \ket{\psi_k}$ with 
$\sum_k |c_k|^2 = 1$.)  By integrating out the stretched horizon states 
$\ket{\tilde{\imath}; k}$ in $\rho_{\rm BH}$, we reproduce the exact 
thermal state, Eq.~(\ref{eq:rho_ext}), for the near exterior states:
\begin{equation}
  \rho_{\rm ext} = \frac{1}{Z} \sum_i e^{-\beta E_i} \ket{i} \bra{i},
\nonumber
\end{equation}
where we have used Eq.~(\ref{eq:stretched}).  The spectrum of Hawking 
radiation, therefore, is exactly thermal, up to gray-body factors which 
arise from (calculable) effects of potential barriers on the outgoing 
quanta.

The evolution of the state described above, which can be summarized by 
Eqs.~(\ref{eq:young},~\ref{eq:old}), implies that the information about 
the initial collapsing matter is encoded mostly in the coefficients $c_k$ 
at an early stage of the black hole evolution, and then in the $d_k$'s and 
$\ket{r_k}$'s (or in the coefficients $d_k$ and $g_{ka}$ if $\ket{r_k}$'s 
are expanded in fixed basis states $\ket{e_a}$ describing the far exterior 
region, $\ket{r_k} = \sum_a g_{ka} \ket{e_a}$).  Finally, after the 
black hole is evaporated, the information is contained in the state 
of final-state Hawking radiation.  We can write this transfer of the 
information schematically as%
\footnote{Note that this description is only schematic.  For example, at 
 an early stage of the evolution, the information about the initial matter 
 may be contained not only in $c_k$'s but also in the deviation of the 
 black hole state from a vacuum state, i.e.\ the appearance of general 
 $\ket{\phi_k}$'s instead of $\ket{\psi_k}$'s in Eq.~(\ref{eq:young}). 
 The following description, however, still gives the correct overall 
 picture for the transfer of the entropy throughout the whole process.}
\begin{equation}
  \ket{M_{\rm init}} 
  \;\rightarrow\; \{ c_k \} 
  \;\rightarrow\; \{ d_k, \ket{r_k} \} 
  \;\rightarrow\; \ket{r_{\rm fin}}.
\label{eq:transfer}
\end{equation}
Since all the information about the initial state is kept, the evolution 
is unitary.  Note that unitarity is preserved here in such a way that the 
entropy of (or the information about) the initial collapsing matter is 
transformed into that of a vacuum and then to that of final-state Hawking 
radiation.  In the intermediate stage of this evolution, entanglement 
between the vacuum and (early) Hawking radiation develops, which is 
transferred to the entanglement among final-state Hawking quanta after 
the evaporation.  This exchange of information between matter/radiation 
and a vacuum makes the black hole formation and evaporation processes 
particularly interesting (and, perhaps, is the main reason why these 
processes are hard to understand).

\subsection{The fate of an infalling object and interior spacetime}
\label{subsec:infalling}

What happens if an object falls into the horizon?  From the viewpoint 
of a distant reference frame, the object will interact with the stretched 
horizon degrees of freedom which have a Planckian temperature---it is 
absorbed into the stretched horizon, and after time (at least) of order 
$M l_{\rm P}^2 \ln(M l_{\rm P})$, its information will be sent back to 
the exterior in the form of Hawking radiation.  On the other hand, general 
relativity tells us that the infalling object should not feel anything 
special at the horizon and must simply fall into the interior.  In 
order to reproduce this picture, the physics of the stretched horizon 
after the object has fallen (which is strongly interacting when viewed 
from a distant reference frame) must be able to be organized into a 
form that allows for the interpretation that the object falls freely 
through the interior spacetime region.

We now discuss how this picture can come out in our framework.  The state 
of the system with a black hole is given by Eq.~(\ref{eq:old}) (or its 
special case, Eq.~(\ref{eq:young})).  Suppose we let an object, which 
was originally located in a far exterior region, fall into the black hole. 
The state of the entire system at the beginning of the fall can then be 
written as
\begin{equation}
  \ket{\Psi_0} = \left( \sum_{k=1}^{e^{\approx {\cal A}/4 l_{\rm P}^2}}\! 
    d_k \ket{\psi_k}\, \ket{r_k} \right) \ket{\chi},
\label{eq:drop-init}
\end{equation}
where $\ket{\chi}$ is the initial state of the object.  The subsequent 
evolution of the system, in particular the falling of the object into 
the black hole, can be described by acting the time evolution operator 
$e^{-iHt}$ on this state $\ket{\Psi_0}$.  In accordance with the 
complementarity hypothesis, we assume that this operator can be organized 
in a way that makes it manifest that the object passes through the horizon 
without being disrupted.  (In the language of Ref.~\cite{Nomura:2011rb}, 
this corresponds to changing the reference frame from a distant to 
an infalling one.)  What is the form of the Hamiltonian $H$ in such 
a description?

Let us denote the Hamiltonian in the standard Unruh-Israel 
description of a black hole written in terms of infalling modes by
\begin{equation}
  H_{\rm UI}\bigl( a_{\Omega,\xi}, a_{\Omega,\xi}^\dagger; 
    c_{\bf p}, c_{\bf p}^\dagger \bigr) 
  \simeq H_{\rm UI,\,Rind}\bigl( a_{\Omega,\xi}, a_{\Omega,\xi}^\dagger \bigr) 
    + H_{\rm UI,\,far}\bigl( c_{\bf p}, c_{\bf p}^\dagger \bigr),
\label{eq:H_UI}
\end{equation}
where $c_{\bf p}^\dagger$/$c_{\bf p}$ are the creation/annihilation 
operators for the modes in the far exterior region, $r \simgt 3M 
l_{\rm P}^2$.  In the last expression, we have separated the Hamiltonian 
into those describing the Rindler ($\approx$~interior $+$ near exterior) 
region, $H_{\rm UI,\,Rind}$, and the far exterior region, $H_{\rm UI,\,far}$, 
by ignoring the terms coupling $a_{\Omega,\xi}$'s and $c_{\bf p}$'s. 
(For simplicity, here we focus on the dynamics of an object falling into 
a black hole of fixed mass $M$, which is a good approximation given the 
timescale of any object to hit the singularity.)  We postulate that the 
dynamics of the infalling object can be well described by the ``infalling 
Hamiltonian''
\begin{equation}
  H = \sum_{k=1}^{e^{\approx {\cal A}/4 l_{\rm P}^2}}\! H^{(k)} P_k;
\qquad
  H^{(k)} = H_{\rm UI}\bigl( a_{\Omega,\xi}^{(k)}, 
    a_{\Omega,\xi}^{(k)\dagger}; c_{\bf p}, c_{\bf p}^\dagger \bigr),
\label{eq:infall-H}
\end{equation}
where $a_{\Omega,\xi}^{(k)}$ and $a_{\Omega,\xi}^{(k)\dagger}$ are given 
by Eq.~(\ref{eq:a_k}) and its conjugate, respectively, and $P_k$ is the 
projection operator defined just below it.  The evolution of the state 
$\ket{\Psi_0}$ is then
\begin{equation}
  \ket{\Psi_0} \rightarrow e^{-iHt} 
    \left( \sum_{k=1}^{e^{\approx {\cal A}/4 l_{\rm P}^2}}\! 
    d_k \ket{\psi_k}\, \ket{r_k} \right) \ket{\chi} 
  = \sum_{k=1}^{e^{\approx {\cal A}/4 l_{\rm P}^2}}\! 
    d_k \left( e^{-i H^{(k)} t} \ket{\psi_k} \ket{r_k} \ket{\chi} \right).
\label{eq:drop-evol}
\end{equation}
We thus find that each ``branch'' of the state, labeled by $k$, evolves 
independently with its own Hamiltonian $H^{(k)}$, all of which, however, 
have the form identical to that of the standard (infalling) Unruh-Israel 
Hamiltonian.  Moreover, since stretched horizon states with different $k$ 
values are orthogonal (see Eq.~(\ref{eq:stretched})), each branch behaves 
as an independent world that does not interfere with others.  This, therefore, 
explicitly realizes the idea suggested in Ref.~\cite{Nomura:2012ex} to 
have a smooth horizon consistently with unitary evolution of a black hole.

We note that the Hamiltonian in Eq.~(\ref{eq:infall-H}) may also be 
written as
\begin{equation}
  H \simeq \left( \sum_{k=1}^{e^{\approx {\cal A}/4 l_{\rm P}^2}}\! 
    H_{\rm UI,\,Rind}\bigl( a_{\Omega,\xi}^{(k)}, 
      a_{\Omega,\xi}^{(k)\dagger} \bigr) \right) 
    + H_{\rm UI,\,far}\bigl( c_{\bf p}, c_{\bf p}^\dagger \bigr),
\label{eq:infall-H-2}
\end{equation}
where we have ignored the terms involving both $a_{\Omega,\xi}^{(k)}$'s 
and $c_{\bf p}$'s, which exist near the boundary between the near and far 
regions, and have used $a_{\Omega,\xi}^{(k)} = a_{\Omega,\xi}^{(k)} P_k$ 
and $\sum_k P_k = 1$ in the first and second terms, respectively.  This 
expression makes it clear that the Hamiltonian in the far exterior region 
has not been modified from the standard form in Eq.~(\ref{eq:H_UI}), as is 
naturally expected.  In fact, if we define the infalling mode operators by
\begin{equation}
  a_{\Omega,\xi} \equiv \sum_k a_{\Omega,\xi}^{(k)},
\qquad
  a_{\Omega,\xi}^\dagger \equiv \sum_k a_{\Omega,\xi}^{(k)\dagger},
\label{eq:modes}
\end{equation}
then we find that all the usual expressions for field theory operators 
go through because of the projection operator $P_k$ involved in 
$a_{\Omega,\xi}^{(k)}$.  For example, the Hamiltonian is given simply 
by $H =  H_{\rm UI}( a_{\Omega,\xi}, a_{\Omega,\xi}^\dagger; c_{\bf p}, 
c_{\bf p}^\dagger)$, and the creation/annihilation operator algebra by 
$[a_{\Omega,\xi}, a_{\Omega',\xi'}^\dagger] = \delta_{\Omega\Omega'} 
\delta_{\xi\xi'}$, $[a_{\Omega,\xi}, a_{\Omega',\xi'}] = 
[a_{\Omega,\xi}^\dagger, a_{\Omega',\xi'}^\dagger] = 0$.

The analysis described above indicates that the stretched horizon degrees 
of freedom provide the states necessary to compose (the $e^{\approx 
{\cal A}/4 l_{\rm P}^2}$ copies of) the second exterior region of the 
Rindler space, and hence the near-horizon eternal black hole geometry. 
In particular, the quantum mechanical structure of a collapse-formed 
black hole (often called a one-sided black hole) after the horizon 
is stabilized to a generic state is that of an eternal (two-sided) 
black hole at the microscopic level.  We can summarize these points 
by the following statement:
\begin{align}
  & \mbox{One-sided black hole with a stretched horizon}
  = \mbox{Two-sided black hole}
\nonumber\\
  &\qquad = \mbox{A ``superposition'' of } e^{\approx 
    \frac{\cal A}{4l_{\rm P}^2}} \mbox{ Unruh-Israel near-horizon quantum 
    field theories}.
\label{eq:slogan}
\end{align}
Here, the quotation marks around the word ``superposition'' indicate that 
the states of the (near-horizon) Unruh-Israel theories may be entangled 
with far exterior states, as in Eq.~(\ref{eq:old}).  We emphasize that 
the correspondence between the collapse-formed and eternal black holes 
discussed here applies at each instant of time (or in a sufficiently short 
time period compared with the timescale for the evolution of the black 
hole).  In particular, the mass of the corresponding (hypothetical) eternal 
black holes must be taken as that of the evolving black hole at each 
moment $M(t)$, not the initial mass $M_0$.

We now argue that the dynamics of the stretched horizon postulated in 
Eq.~(\ref{eq:infall-H}) implies that an infalling observer finds that 
the horizon is smooth (no drama) with a probability of $1$.  Suppose 
the initial state before the infall was given by $\ket{\Psi_0}$ in 
Eq.~(\ref{eq:drop-init}), and that the observer does not interact strongly 
with the system entangled with the black hole (early Hawking radiation) 
throughout the falling.  Now, the state $\ket{r_k}$ is in general 
a superposition of decohered classical states $\ket{r^{\rm cl}_m}$, 
$\ket{r_k} = \sum_m U_{km} \ket{r^{\rm cl}_m}$ where $U_{km}$ is 
a unitary matrix, and the observer finds himself/herself to live 
in one of these worlds with the (collapsed) state given by
\begin{equation}
  \ket{\Psi_0^{(m)}} = \frac{1}{\sqrt{\sum_{k'=1}^{e^{\approx {\cal A}/4 
    l_{\rm P}^2}}\!  |d_{k'} U_{k'm}|^2}} 
  \sum_{k=1}^{e^{\approx {\cal A}/4 l_{\rm P}^2}}\! 
    d_k\, U_{km} \ket{\psi_k} \ket{r^{\rm cl}_m} \ket{\chi}.
\label{eq:drop-m}
\end{equation}
According to Eq.~(\ref{eq:drop-evol}), when this observer interacts with 
the black hole state, he/she will find himself/herself to be in a particular 
vacuum $\ket{\psi_k}$ with probability $|d_k U_{km}|^2/\sum_{k'} |d_{k'} 
U_{k'm}|^2$ (i.e.\ the measurement basis is $\ket{\psi_k}$), but all 
of these vacua lead to the same semi-classical physics dictated by the 
Hamiltonian $H_{\rm UI}$, i.e.\ general relativity.  The fact that the 
observer finds a smooth horizon with a probability of $1$ does not change 
even if he/she performs an arbitrary measurement on early Hawking radiation 
before jumping into the black hole.  Without loss of generality, we may 
assume that the outcome of the measurement was $\sum_k V_{nk} \ket{r_k}$, 
where $V_{nk}$ is an arbitrary unitary matrix.  By repeating the same 
argument as above with $U_{km}$ replaced by $V^\dagger_{kn}$, we find 
that the observer does not see anything special at the horizon.%
\footnote{On the other hand, if a falling observer could directly measure 
 states entangled with the horizon and then enter to it right after 
 (i.e.\ before the state of the black hole changes), then he/she may 
 see a firewall.  (It is not clear if such a measurement can indeed 
 be performed; it is possible that there is some dynamical, or perhaps 
 computational~\cite{Harlow:2013tf}, obstacle to it.)  This, however, 
 does not violate the equivalence principle, since the same argument 
 applies to any surface in a low curvature region, i.e.\ there is nothing 
 special about the black hole horizon.}

How does the analysis described above evade arguments for firewalls? 
As discussed in Ref.~\cite{Nomura:2012ex}, the entropy argument based on 
the strong subadditivity relation~\cite{Almheiri:2012rt} is avoided because 
the entropies relevant for discussing unitarity are different from those 
for the smoothness of the horizon.  Specifically, unitarity requires 
the von~Neumann entropies of subsystems consisting of $A$, $B$, and $R$ 
(the stretched horizon, near horizon region, and far exterior region, 
respectively) calculated using the entire quantum state $e^{-iHt}\, 
\ket{\Psi_0}$ to satisfy $S_{BR} < S_R$.  On the other hand, the 
smoothness of the horizon requires $\tilde{S}^{(k)}_{AB} \approx 0$ 
(for all $k$), where $\tilde{S}^{(k)}_X$ is the von~Neumann entropy 
of subsystem $X$ calculated using the ``branch state'' $e^{-iH^{(k)}t}\, 
U_{km} \ket{\psi_k} \ket{r^{\rm cl}_m} \ket{\chi}/|U_{km}|$ without 
summation over $k$.  These two relations are {\it not} incompatible 
with each other, unlike the case if they were both calculated using 
the same quantum state.

The typicality argument of Ref.~\cite{Marolf:2013dba} is also avoided 
because the black hole vacuum states $\ket{\psi_k}$ cannot all be 
transformed into eigenstates of the number operator for a near exterior 
mode, $\hat{b}^\dagger \hat{b}$, by performing a unitary rotation in 
the space spanned by $\ket{\psi_k}$, ${\cal H}_\psi$.  (For a similar 
discussion, see Ref.~\cite{Nomura:2013gna}.)  Specifically, by expanding 
near horizon states $\ket{i}$ by the $\hat{b}^\dagger \hat{b}$ eigenstates 
$\ket{e_j}$ as $\ket{i} = \sum_j c^i_j \ket{e_j}$, the black hole 
vacuum states become 
\begin{equation}
  \ket{\psi_k} = \frac{1}{\sqrt{Z}} \sum_j \ket{e_j} \left( \sum_i 
    e^{-\frac{\beta}{2}E_i} c^i_j\, \ket{\tilde{\imath}; k} \right).
\label{eq:phi_bb-expand}
\end{equation}
We find that because of the index $k$ in the stretched horizon states, 
we cannot find a basis change in ${\cal H}_\psi$ that makes all the 
$\ket{\psi_k}$'s $\hat{b}^\dagger \hat{b}$ eigenstates, and hence there 
is no reason to expect that typical black hole states have firewalls. 
Indeed, by calculating the average number of high energy quanta in 
an infalling frame (i.e.\ quanta with $\Omega \gg 1/M l_{\rm P}^2$), 
we obtain
\begin{equation}
  \bar{N} \equiv \frac{{\rm Tr}_{{\cal H}_\psi} 
    \hat{N}_a}{{\rm Tr}_{{\cal H}_\psi} {\bf 1}} 
  = \frac{\sum_{k=1}^{e^{\approx {\cal A}/4 l_{\rm P}^2}} \bra{\psi_k} 
    \hat{N}_a \ket{\psi_k}}{e^{\approx {\cal A}/4 l_{\rm P}^2}} 
  \approx 0,
\label{eq:average}
\end{equation}
because of Eq.~(\ref{eq:inf_vac_k}), where $\hat{N}_a$ is the number 
operator for the quanta with the indices $\Omega$ and $\xi$
\begin{equation}
  \hat{N}_a = \sum_{k=1}^{e^{\approx {\cal A}/4 l_{\rm P}^2}}\! 
    a_{\Omega,\xi}^{(k)\dagger} a_{\Omega,\xi}^{(k)}.
\end{equation}
We find that typical black hole states do {\it not} have firewalls.  (In 
fact, no black hole state has a firewall.)

\subsection{Complementarity as a unitary reference frame change}
\label{subsec:compl}

We have seen that the dynamics of the stretched horizon is such that 
it can produce $e^{\approx {\cal A}/4 l_{\rm P}^2}$ copies of quantum 
field theories describing physics in the interior region of a black 
hole.  Is it then possible to organize the description of the entire 
system in a way that makes manifest the local nature of the interior 
spacetime region while keeping unitarity at the full quantum level? 
Here, following Ref.~\cite{Nomura:2011rb}, we consider that such an 
``infalling description'' is obtained by performing a unitary complementarity 
transformation on a distant description, which corresponds to changing 
the (local Lorentz) reference frame from a distant one to an infalling 
one.  What does this description look like?

If the complementarity transformation is indeed unitary, the $e^{\approx 
{\cal A}/4 l_{\rm P}^2}$ states $\ket{\psi_k}$ must be transformed into 
$e^{\approx {\cal A}/4 l_{\rm P}^2}$ different states which must all look 
locally like Minkowski vacuum states.  In particular, this implies that 
in the limit that the black hole is large ${\cal A} \rightarrow \infty$, 
i.e.\ in the limit that the horizon under consideration is a Rindler 
horizon, there are infinitely many Minkowski vacuum states labeled by 
$k=1,\cdots,e^{\approx {\cal A}/4 l_{\rm P}^2}=\infty$.  This seems 
to contradict our experience that we can do physics without knowing 
which of the Minkowski vacua we live in.  Isn't the Minkowski vacuum 
unique, e.g., in QED?  Otherwise, we do not seem to be able to do 
any physics without having the (infinite amount of) information on 
the Minkowski vacua.

The answers to these questions arise by noticing that when described 
in an infalling reference frame, the black hole spacetime is Minkowski 
vacuum-like only locally, and the nonzero curvature effect can lead 
to a ``horizon'' (as viewed from the infalling frame, not the original 
one as viewed from a distant frame) at a finite spatial distance, 
a distance of order the Schwarzschild radius away from the origin of 
the reference frame (see Ref.~\cite{Nomura:2013nya} for a related 
discussion).  This makes it possible that the $e^{\approx {\cal A}/4 
l_{\rm P}^2}$ different states in the infalling description correspond 
to different configurations of the degrees of freedom on this ``horizon,'' 
and that physics in the interior region is given by the same local 
Hamiltonian in all these $e^{\approx {\cal A}/4 l_{\rm P}^2}$ states, 
consistent with the postulate in Eq.~(\ref{eq:infall-H}).  In order 
for an observer to know which $\ket{\psi_k}$ vacuum he/she is in, 
he/she needs to probe (intrinsically quantum gravitational) degrees 
of freedom on his/her ``horizon''; the effect he/she can probe locally 
away from the ``horizon'' is expected to be suppressed exponentially 
in ${\cal A}/l_{\rm P}^2$.  In the true Minkowski space limit of 
${\cal A} \rightarrow \infty$, this ``horizon'' is located only at 
spatial infinity, which no physical observer can access.  The uniqueness 
of the Minkowski vacuum (for the purpose of describing local physics) 
is recovered in this way.

Finally, we comment on unitary equivalence relations between subsystems 
in distant and infalling reference frames.  When viewed from a distant 
reference frame, the information about an object absorbed into the 
stretched horizon stays there for, at least, a time of order $M l_{\rm P}^2 
\ln(M l_{\rm P})$, which will later be sent back to the exterior in 
Hawking radiation.  On the other hand, when we describe the same object 
in an infalling reference frame (whose origin closely follows the 
trajectory of the object), the description of the object being in the 
interior spacetime is available only until the object or the origin of 
the reference frame hits the singularity a time of order $M l_{\rm P}^2$ 
after the fall; after this time the system is described only by ``singularity 
states''~\cite{Nomura:2011rb}:\ intrinsically quantum gravitational states 
that do not allow for a spacetime interpretation.  (The infalling reference 
frame is obtained by performing a boost transformation on a distant 
reference frame at some time; see Ref.~\cite{Nomura:2012cx} for more 
detailed discussions on this prescription.)  This implies that when 
the complementary transformation is defined as the relation between 
the two descriptions at the same proper time of the origin of their 
respective reference frames, then the interior spacetime can correspond 
only to the stretched horizon degrees of freedom while Hawking radiation 
only to the singularity degrees of freedom:
\begin{equation}
\begin{array}{ccc}
  \mbox{infalling}   &         & \mbox{distant} \\ \hline
  \mbox{interior}    & \subset & \mbox{stretched horizon} \\
  \mbox{singularity} & \supset & \mbox{Hawking radiation}.
\end{array}
\end{equation}
In particular, this seems to prevent the possibility of ``gravity/gravity'' 
duality in black hole physics, in which information in Hawking radiation 
is directly mapped to that in the interior spacetime at the same proper 
time of the two (distant and infalling) reference frames.

We may even go further.  As we have seen, and is particularly clear in 
Eq.~(\ref{eq:modes}), an object traveling in the interior spacetime is 
described by isomorphic excitations of the $e^{{\cal A}/4 l_{\rm P}^2}$ 
vacuum states $\ket{\psi_k}$, without their coefficients $d_k$ in the 
state of the entire system changing throughout the falling process. 
Namely, when viewed from a distance, the information about such an object 
is contained in the deviations of the black hole states from the vacuum 
states, $\ket{\phi_k} \neq \ket{\psi_k}$, and not in $d_k$.  In fact, 
by the time the information about the fallen object becomes distributed 
into the coefficients $d_k$ in the distant picture, the object in 
the corresponding interior picture has already been absorbed into the 
``horizon'' surrounding it.  Note that the distance to the ``horizon'' 
from the origin of the infalling reference frame becomes smaller as the 
singularity is approached (and becomes less than $l_*$ at some point near 
the singularity) because of the increase of the curvature effect there. 
This implies that there is no complementary description of the information 
scrambling, or evaporation, process in the interior picture in the regime 
where the semi-classical spacetime description is applicable.

\section{More General Spacetimes}
\label{sec:general}

In this section, we discuss how the picture developed so far in black 
hole physics can be extended to more general spacetimes.  The structure 
of the section is such that the discussion becomes more conjectural as 
it progresses.  We first consider a relatively straightforward application 
of the dynamics of the stretched horizon we have learned in black hole 
physics to de~Sitter space.  We then discuss possible implications of 
our picture for the general structure of Hilbert space in quantum gravity.

\subsection{de~Sitter Space}
\label{subsec:de-Sitter}

Consider de~Sitter space with the Hubble parameter $H$.  The de~Sitter 
horizon is located at $r = 1/H$, where $r$ is the radial coordinate of the 
static coordinate system $(t, r, \theta, \phi)$.  The stretched horizon is 
located where the local Gibbons-Hawking temperature~\cite{Gibbons:1977mu}
\begin{equation}
  T(r) = \frac{H/2\pi}{\sqrt{1-H^2 r^2}},
\label{eq:T_dS}
\end{equation}
becomes of order the fundamental, or string, scale $1/l_*$:\ $T(r_*) = 
1/2\pi l_*$ (where the factor of $2\pi$ is chosen so that the local proper 
acceleration of a fixed spatial coordinate point at the stretched horizon 
is $1/l_*$), i.e.\
\begin{equation}
  r_* = \frac{1}{H} - \frac{1}{2} H l_*^2.
\label{eq:r_*}
\end{equation}
Following the discussion in the black hole case, we consider that the 
stretched horizon degrees of freedom are organized into the states labeled 
by $\tilde{\imath}$ and $k$:
\begin{equation}
  \ket{\tilde{\imath}; k}
\quad\mbox{with}\quad
  \inner{\tilde{\imath}; k}{\tilde{\imath}'; k'} 
  = \delta_{\tilde{\imath}\tilde{\imath}'} \delta_{kk'},
\label{eq:dS-stretched}
\end{equation}
and that the $\tilde{\imath}$ index is entangled with the states in the 
interior of the stretched de~Sitter horizon $\ket{i}$ (corresponding to 
the near-horizon states outside the stretched Schwarzschild horizon) as
\begin{equation}
  \ket{\psi_k} = \frac{1}{\sqrt{\sum_{i'} e^{-\beta E_{i'}}}} \sum_i 
    e^{-\frac{\beta}{2}E_i} \ket{i} \ket{\tilde{\imath}; k},
\label{eq:dS-phi_k}
\end{equation}
in the de~Sitter vacuum states.  Here, $E_i$ is the energy of $\ket{i}$ 
measured at the origin $r = 0$, and $\beta = 2\pi/H$ the inverse 
Gibbons-Hawking temperature; the index $k$ runs over
\begin{equation}
  k = 1, \cdots, e^{\approx \frac{{\cal A}}{4 l_{\rm P}^2}},
\label{eq:dS-k}
\end{equation}
where ${\cal A} = 4\pi/H^2$ is the area of the (stretched) de~Sitter horizon.

We assume, given the absence of evidence otherwise, that the analysis 
of the stretched horizon in the black hole case can be straightforwardly 
adapted to de~Sitter space (with the obvious interchange of the ``interior'' 
and ``exterior'':\ the region outside the de~Sitter horizon corresponds 
to the region inside the Schwarzschild horizon).  When viewed from 
a reference frame associated with the static coordinates, the entropy 
of the system is saturated at the leading order in $l_{\rm P}^2/{\cal A}$ 
by the logarithm of the possible number of vacuum states, $\ket{\psi_k}$:
\begin{equation}
  S_{\rm dS} \approx \ln{\rm dim} {\cal H}_\psi 
  \approx \frac{\cal A}{4 l_{\rm P}^2},
\label{eq:A_dS}
\end{equation}
where ${\cal H}_\psi$ is the Hilbert space spanned by $\ket{\psi_k}$. 
The number of possible states the interior region can be in (without 
forming a black hole, which would change the horizon structure) as 
well as the number of possible states for the near exterior region 
are expected to contribute only negligibly, by an amount smaller 
powers in ${\cal A}/l_{\rm P}^2$.  (The states in which there is matter 
in the near exterior region can be obtained by acting creation operators 
$a_{\Omega,\xi}^{(k)\dagger}$ on $\ket{\psi_k}$, constructed analogously 
to the black hole case.)  When a reference frame change corresponding to 
a shift of the origin of the reference frame is performed, a spacetime 
region outside the original horizon can be reconstructed, with the 
excitations in the near exterior region corresponding to perturbations 
of $\ket{\psi_k}$ by $a_{\Omega,\xi}^{(k)\dagger}$'s while those in 
the far exterior region involving the index $k$.  With a succession 
of such reference frame changes, the (approximate) picture of 
global de~Sitter space may be obtained, which, however, grossly 
overcounts the number of degrees of freedom if the horizon degrees 
of freedom (e.g.\ in each Hubble volume) are also included in the 
description~\cite{Nomura:2011dt,Nomura:2011rb}.

Following the same analysis as in the black hole case, we find that 
the de~Sitter horizon is smooth:\ an object that hits the horizon 
can be thought of as going to space outside the horizon.  The information 
about the object that goes outside will be stored in the general state 
of the form
\begin{equation}
  \ket{\phi_k} = \frac{1}{\sum_{i',\tilde{\jmath}'} |f_{i,\tilde{\jmath}}|^2} 
    \sum_{i,\tilde{\jmath}} f_{i\tilde{\jmath}}\, 
    \ket{i} \ket{\tilde{\jmath}; k},
\label{eq:dS-gen}
\end{equation}
constructed purely from the interior and the stretched horizon degrees 
of freedom.  Such information may thus be recovered later.  This information 
recovery may not necessarily be in the form of Hawking radiation if the 
system evolves, for example, into Minkowski space or another de~Sitter 
space with a smaller vacuum energy.  Indeed, this is believed to have 
happened to density fluctuations generated in the early inflationary 
phase in our universe~\cite{Hawking:1982cz}.

In the limit $H \rightarrow 0$ in which the de~Sitter space approaches 
Minkowski space, the number of vacuum states becomes infinity
\begin{equation}
  {\rm dim}\,{\cal H}_\psi \rightarrow \infty.
\label{eq:H-Minkoski}
\end{equation}
Since the horizon is located at spatial infinity in this limit, probing 
the structure of (infinitely many) Minkowski vacua will require access 
to the horizon at infinity.  This is the same picture as the one arrived 
at in Section~\ref{subsec:compl} by taking the large mass limit of 
a black hole.

\subsection{Hilbert space for quantum gravity and the entropy bound}
\label{subsec:bound}

What does the picture developed so far imply for more general spacetimes 
in quantum gravity?  Following Refs.~\cite{Nomura:2011rb,Nomura:2013nya}, 
here we consider that the Hilbert space for quantum gravity can be 
organized such that the system is viewed from a freely falling (local 
Lorentz) reference frame.  (This corresponds to partially fixing large 
gauge redundancies in full quantum gravity.)  We will be agnostic about 
its detailed implementation, e.g., whether a null or spacelike quantization 
is used inside horizons as viewed from the reference frame.

The evolution of a system is described by giving a quantum state at each 
time $\tau$, taken as the proper time measured at the spatial origin 
$p(\tau)$ of the reference frame.  These states are in general superpositions 
of component states that represent configurations on well-defined 
semi-classical (``equal-time'') spacetime hypersurfaces.  (More precisely, 
the state of the system may also contain ``singularity states'' 
that do not allow for a spacetime interpretation; these states are 
relevant when $p(\tau)$ hits a singularity.)  Now, we can group these 
component states, which span a Hilbert space ${\cal H}$, into classes 
${\cal H}_{\partial{\cal M}}$ that represent all possible physical 
configurations in all possible spacetime hypersurfaces that share the 
same boundary $\partial{\cal M}$:
\begin{equation}
  {\cal H} = \bigoplus_{\partial{\cal M}} {\cal H}_{\partial{\cal M}}.
\label{eq:H-H_M}
\end{equation}
Our interest is in how many independent quantum states there are in 
${\cal H}_{\partial{\cal M}}$ with a fixed $\partial{\cal M}$, and what 
the entropy associated with them (i.e.\ the logarithm of that number) 
corresponds to.  Note that quantum states we discuss here 
are those for the entire system; in particular, they include states 
for the boundary degrees of freedom.

Let us consider a fixed spacetime, corresponding roughly to some 
fixed $\partial{\cal M}$.  The origin of the reference frame $p(\tau)$ 
is then typically surrounded by horizons where Planckian physics becomes 
important, which we take as our boundaries.  (Note that the horizons 
may be located at spatial infinity as in the case of Minkowski space.) 
For example, if $p(\tau)$ is located outside the Schwarzschild horizon 
in a de~Sitter space with a black hole, then $p(\tau)$ will ``see'' the 
black hole horizon in some directions and the de~Sitter horizon in the 
others.  Specifically, suppose that the quantization surface at time 
$\tau$ is parameterized by coordinates $(\lambda,\theta,\phi)$ which, 
for infinitesimal $\lambda$, are spherical coordinates for the locally 
inertial reference frame of $p(\tau)$.  Then, a radial axis with constant 
$(\theta,\phi)$ will hit either the (stretched) black hole or de~Sitter 
horizon, depending on the values of $(\theta,\phi)$.  Now, the area 
of these horizons are
\begin{equation}
  {\cal A}_{\rm BH} \sim 16\pi M^2 l_{\rm P}^4,
\qquad
  {\cal A}_{\rm dS} \sim \frac{4\pi}{H^2},
\label{eq:horizon-areas}
\end{equation}
where $M$ and $H$ are the mass of the black hole and the Hubble parameter, 
respectively, and we have assumed $M l_{\rm P}^2 \ll 1/H$.  What is the 
entropy of this system, i.e.\ the logarithm of the number of possible 
independent quantum states that have the same horizon/boundary structure?

The covariant entropy bound~\cite{Bousso:1999xy} suggests that the 
(fine-grained) entropy of this spacetime, more precisely the logarithm 
of the number of independent quantum states consistent with the boundary 
structure described above, is given by
\begin{equation}
  S \approx \frac{1}{4 l_{\rm P}^2} \left( {\cal A}_{\rm BH} 
    + {\cal A}_{\rm dS} \right),
\label{eq:S_cov-SdS}
\end{equation}
where we have assumed that the bound is indeed saturated.  What does 
this entropy count?  Performing an analysis similar to the one in 
Ref.~\cite{'tHooft:1993gx}, we expect that the entropy arising from the 
``interior'' configurations, i.e.\ configurations between the Schwarzschild 
and de~Sitter horizons, is negligible.  (Note that we should not count 
the thermal entropy of Hawking radiation for this purpose; it could arise 
even if the entire state, which includes the horizon degrees of freedom, 
were unique, i.e.\ $S = 0$, as long as the interior and horizon degrees 
of freedom are entangled.)  The leading contributions then must come 
from the horizons as we discussed in Sections~\ref{sec:BH-entropy} 
and \ref{subsec:de-Sitter}.  These entropies are the entropies of a 
vacuum---when a reference frame change is performed, the relevant 
degrees of freedom are mapped into those in the horizons of the new 
reference frame, which dictate how many interior quantum field theories 
(which all look identical) can be realized at the full quantum gravity 
level.  For example, if $p(\tau)$ is transformed to be in the interior 
of the black hole, then $S$ is realized mostly as the logarithm of 
possible configurations of the degrees of freedom on the ``horizon'' 
discussed in Section~\ref{subsec:compl}.  When $p(\tau)$ is transformed 
to be in the exterior of the de~Sitter horizon, $S$ mostly counts the 
logarithm of possible configurations of the degrees of freedom on the 
new de~Sitter horizon surrounding the new, transformed $p(\tau)$.

More generally, the fine-grained entropy of the system---the logarithm 
of the number of independent states in ${\cal H}_{\partial{\cal M}}$---is 
given by the area of the Planckian boundaries surrounding $p(\tau)$ 
(which may exist at spatial infinity):%
\footnote{The general definition of the structures of two boundaries 
 being the same is not obvious; one possible definition is to require 
 that they explicitly have the same induced metric in terms of the 
 coordinates used in defining the theory, e.g.\ $(\lambda,\theta,\phi)$ 
 used above; see Ref.~\cite{Nomura:2013nya} for a discussion on this 
 point.}
\begin{equation}
  S \approx \frac{1}{4 l_{\rm P}^2} 
    \int_{(\theta,\phi)} d{\cal A},
\label{eq:S_cov}
\end{equation}
where we have assumed that the spacelike projection theorem of 
Ref.~\cite{Bousso:1999xy} applies if a spacelike quantization is 
employed.  There is, however, one important caveat.  Suppose we follow 
a radial axis, parameterized by $\lambda$, in the direction of constant 
$(\theta,\phi)$ and find that it hits an apparent horizon before hitting 
a Planckian boundary such as the stretched black hole or de~Sitter 
horizon.  Here, the apparent horizon is defined locally as a surface 
on which the expansion of the past-directed outgoing light rays, emitted 
from (a portion of) the constant $\lambda$ surface, first crosses 
zero. (An example is given by a surface at $r = (1/a(t) l_{\rm P}) 
\sqrt{3/8\pi \rho(t)}$ in the Friedmann-Robertson-Walker universe, where 
we have assumed that $p(\tau)$ is comoving at $r=0$, and $a(t)$ and 
$\rho(t)$ are the scale factor and energy density, respectively.) 
If this happens, then the surface used to bound the entropy must be 
replaced with the apparent horizon.  Specifically, the expression 
in Eq.~(\ref{eq:S_cov}) must be modified to~\cite{Nomura:2013nya}
\begin{equation}
  S \approx \frac{1}{4 l_{\rm P}^2} 
    \int_{(\theta,\phi)} d{\cal A}(\lambda_{\rm H}(\theta,\phi));
\qquad
  \lambda_{\rm H}(\theta,\phi) = {\rm min}\left\{ \lambda_{\rm P}(\theta,\phi), 
    \lambda_{\rm app}(\theta,\phi) \right\},
\label{eq:S_cov-2}
\end{equation}
where $d{\cal A}(\lambda_{\rm H}(\theta,\phi))$ is an area element 
for a surface $\lambda = \lambda_{\rm H}(\theta,\phi)$, and 
$\lambda_{\rm P}(\theta,\phi)$ and $\lambda_{\rm app}(\theta,\phi)$ 
are the radial coordinate distances from $p(\tau)$ to the nearest 
Planckian and apparent horizons, respectively. 

We may decompose the expression for $S$ in Eq.~(\ref{eq:S_cov-2}) into 
two pieces as
\begin{equation}
  S \approx S_{\rm P} + S_{\rm app},
\label{eq:S_cov-decomp}
\end{equation}
where
\begin{equation}
  S_{\rm P} \approx \frac{1}{4 l_{\rm P}^2} 
    \int_{(\theta,\phi)_{\rm P}} d{\cal A}(\lambda_{\rm P}(\theta,\phi)),
\qquad
  S_{\rm app} \approx \frac{1}{4 l_{\rm P}^2} 
    \int_{(\theta,\phi)_{\rm app}} d{\cal A}(\lambda_{\rm app}(\theta,\phi)),
\label{eq:S_P-S_app}
\end{equation}
with $(\theta,\phi)_{\rm P}$ and $(\theta,\phi)_{\rm app}$ indicating 
the values of $(\theta,\phi)$ in which $\lambda_{\rm P}(\theta,\phi)$ is 
smaller and larger than $\lambda_{\rm app}(\theta,\phi)$, respectively. 
We have seen that, when a reference frame change is performed, $S_{\rm P}$ 
can be viewed as the entropy of a vacuum.  On the other hand, as seen 
in Ref.~\cite{Bousso:2010pm}, $S_{\rm app}$ can be saturated by the 
contribution from quantum field theory degrees of freedom, i.e.\ the 
matter/radiation contribution from the region outside the apparent 
horizon.  We therefore find that typically $S_{\rm P}$ comes dominantly 
from the entropy of a vacuum while $S_{\rm app}$ can come either from 
a vacuum or matter/radiation contribution:
\begin{equation}
  S_{\rm P} \approx S_{\rm P,\,vac} \gg S_{\rm P,\,mat},
\qquad
  S_{\rm app} \approx S_{\rm app,\,vac} \mbox{ or } S_{\rm app,\,mat},
\label{eq:S-contr}
\end{equation}
where $S_{\rm X,\,vac}$ and $S_{\rm X,\,mat}$ represent the vacuum and 
matter/radiation contributions to $S_X$.  In any event, the fine-grained 
entropy arises in general from both vacuum and matter/radiation 
contributions:
\begin{equation}
  S = S_{\rm vac} + S_{\rm mat},
\label{eq:S-final}
\end{equation}
where $S_{\rm vac} = S_{\rm P,\,vac} + S_{\rm app,\,vac}$ and $S_{\rm mat} 
= S_{\rm P,\,mat} + S_{\rm app,\,mat}$, and only the $S_{\rm mat}$ 
contribution is explicitly included as the dynamical degrees of freedom 
in quantum field theories.  With the contribution from a vacuum included, 
we conjecture that the covariant entropy bound is indeed saturated for 
all ${\cal H}_{\partial{\cal M}}$
\begin{equation}
  S = S_{\rm vac} + S_{\rm mat} \approx \frac{1}{4 l_{\rm P}^2} 
    \int_{(\theta,\phi)} d{\cal A}(\lambda_{\rm H}(\theta,\phi)),
\label{eq:S-saturated}
\end{equation}
which indeed seems to be the case in all the systems we have considered.

\section*{Acknowledgments}

This work was supported in part by the Director, Office of Science, Office 
of High Energy and Nuclear Physics, of the US Department of Energy under 
Contract DE-AC02-05CH11231, and in part by the National Science Foundation 
under grant PHY-1214644.

\appendix

\section{Hilbert Space for the Evolution of a Black Hole}
\label{app:Hilbert}

Here we present the Hilbert space relevant for describing the formation 
and evaporation of a black hole (Schwarzschild black hole in 4-dimensional 
spacetime) from a distant reference frame.%
\footnote{The Hilbert space described here is similar to that in 
 Ref.~\cite{Nomura:2013gna}.  A major difference is the structure of 
 the stretched horizon degrees of freedom, denoted by $A$ here and by 
 $\tilde{B}$ in Ref.~\cite{Nomura:2013gna}.  In particular, the dimension 
 of the Hilbert space factor ${\cal H}_{A(M)}$ here is much larger 
 than $e^{\approx {\cal A}/4 l_{\rm P}^2}$ (specified by both the 
 indices $\tilde{\imath}$ and $k$; see Section~\ref{sec:BH-entropy}), 
 while that of ${\cal H}_{\tilde{B}(M)}$ in Ref.~\cite{Nomura:2013gna} 
 is $e^{\approx {\cal A}/4 l_{\rm P}^2}$.}
We first consider a system with a black hole {\it of fixed mass $M$} and 
decompose it into three subsystems:
\begin{itemize}
\item[$A$:]
the degrees of freedom associated with the stretched horizon;
\item[$C$:]
the degrees of freedom associated with the spacetime region close to, 
but outside, the stretched horizon, e.g.\ $r \simlt 3M l_{\rm P}^2$;
\item[$R$:]
the rest of the system (which may contain Hawking radiation emitted 
earlier).
\end{itemize}
Among all the possible quantum states for the $C$ degrees of freedom, 
some are strongly entangled with the states representing $A$.  We call 
the set of these quantum states $B$:
\begin{itemize}
\item[$B$:]
the quantum states representing the states for the $C$ degrees of freedom 
that are strongly entangled with the degrees of freedom described by $A$.
\end{itemize}
(The identification of $B$ depends on the possible existence of matter 
in the region $2M l_{\rm P}^2 < r \simlt 3M l_{\rm P}^2$.  If there is 
extra matter beyond the black hole in $r \simlt 3M l_{\rm P}^2$, it changes 
the identification of the $B$ states in the Hilbert space for the $C$ 
degrees of freedom.)  Below we will ignore the center-of-mass drift and 
spontaneous spin-up of a black hole~\cite{Page:1979tc,Nomura:2012cx}, 
which give only minor effects on the dynamics.  Including these effects, 
however, is straightforward---we simply have to add indices for the 
center-of-mass location and angular momentum of the black hole to 
the states.

We have now divided the system with a black hole of fixed mass $M$ into 
three subsystems $A$, $C$, and $R$.  Since the black hole mass varies 
with time, however, the Hilbert space in which the state of the entire 
system evolves unitarily must take the form
\begin{equation}
  {\cal H} = \bigoplus_M \left( {\cal H}_{A(M)} \otimes 
    \left\{ {\cal H}_{B(M)} \oplus {\cal H}_{C(M)-B(M)} \right\} 
    \otimes {\cal H}_{R(M)} \right) 
  \equiv \bigoplus_M {\cal H}_M,
\label{eq:Hilbert}
\end{equation}
where we have explicitly shown the $M$ dependence of $A$, $B$, $C$, and 
$R$.  Here, ${\cal H}_{C(M)-B(M)}$ is the Hilbert space spanned by the 
states for the $C$ degrees of freedom orthogonal to $B$ (i.e.\ not entangled 
with $A$), and we define ${\cal H}_0$ to be the Hilbert space for the 
system without a black hole.  As the black hole evolves, the state of 
the system moves between different ${\cal H}_M$'s; for example, a state 
that is an element of ${\cal H}_{M_1}$ with some $M_1$ will later become 
an element of ${\cal H}_{M_2}$ with $M_2 < M_1$ (more precisely, a 
superposition of elements in various ${\cal H}_{M_2}$'s).  In this language, 
$\ket{\psi_k(M)}$ in Eqs.~(\ref{eq:young},~\ref{eq:old}) are elements of 
${\cal H}_{A(M)} \otimes {\cal H}_{B(M)}$, while $\ket{r}$ and $\ket{r_k}$ 
are those of ${\cal H}_{R(M)}$; $\ket{r_{\rm fin}}$, representing 
final-state Hawking radiation, is an element of ${\cal H}_0$.

In our present picture, the Hilbert space factor for the stretched horizon 
degrees of freedom can be decomposed as
\begin{equation}
  {\cal H}_{A(M)} = \bigoplus_{k=1}^{e^{\approx {\cal A}/4 l_{\rm P}^2}} 
    {\cal H}^{(k)}_{A(M)},
\label{eq:app-HA-decom}
\end{equation}
where ${\cal A} = 16\pi M^2 l_{\rm P}^4$, and ${\cal H}^{(k)}_{A(M)}$ 
is the Hilbert space factor of which the stretched horizon states 
$\ket{\tilde{\imath}; k}$ are the elements.  Plugging this expression 
into Eq.~(\ref{eq:Hilbert}), we find that the Hilbert space for the 
system with a black hole of mass $M$ takes the form
\begin{equation}
  {\cal H}_M = \bigoplus_{k=1}^{e^{\approx {\cal A}/4 l_{\rm P}^2}} 
    \left( {\cal H}^{(k)}_{A(M)} \otimes 
    \left\{ {\cal H}_{B(M)} \oplus {\cal H}_{C(M)-B(M)} \right\} 
    \otimes {\cal H}_{R(M)} \right) 
  \equiv \bigoplus_{k=1}^{e^{\approx {\cal A}/4 l_{\rm P}^2}} 
    {\cal H}^{(k)}_M.
\label{eq:app-HM-decom}
\end{equation}
We find that ${\cal H}_M$ consists of $e^{\approx {\cal A}/4 l_{\rm P}^2}$ 
Hilbert subspaces ${\cal H}^{(k)}_M$, each of which takes the form 
identical to the Hilbert space of quantum field theory on a fixed classical 
spacetime background with a black hole of mass $M$.

\end{document}